\documentclass[trackchanges]{aastex701}
\usepackage{tablefootnote}
\usepackage{soulutf8}

\defcitealias{Seligman25}{Seligman+ 25}
\defcitealias{Bolin25}{Bolin+ 26}
\defcitealias{Kareta2025}{Kareta+ 25}
\defcitealias{Beniyama2025}{Beniyama 25}
\defcitealias{Santana-Ros2025}{Santana-Ros+ 25}

\begin{document}

\title[Inbound Gas and Dust Evolution of 3I/ATLAS]{The Inbound Gas and Dust Evolution of Interstellar Comet 3I/ATLAS from Optical Spectroscopy and Multi-band Photometry\footnote{Based on observations made with the Southern African Large Telescope (SALT).}}

\author[orcid=0000-0002-0143-9440,gname=T.,sname='Santana-Ros']{T. Santana-Ros}
\affiliation{Departamento de F\'{\i}sica, Ingenier\'{\i}a de Sistemas y Teor\'{\i}a de la Se\~{n}al, Universidad de Alicante, Carr. San Vicente del Raspeig, s/n, 03690 San Vicente del Raspeig, Alicante, Spain}
\affiliation{Institut de Ci\`encies del Cosmos (ICCUB), Universitat de Barcelona (UB), c. Mart\'i Franqu\`es, 1, 08028 Barcelona, Catalonia, Spain}
\email[show]{tsantanaros@icc.ub.edu}

\author[orcid=0000-0003-3425-5178,gname=A., sname=Sergeyev]{A. Sergeyev}
\affiliation{V.N. Karazin Kharkiv National University, Sumska 35, Kharkiv, 61022, Ukraine}
\affiliation{Universit\'e C{\^o}te d'Azur, Observatoire de la C{\^o}te d'Azur, CNRS, Laboratoire Lagrange, France}
\email{alexey.v.sergeyev@gmail.com}

\author[orcid=0000-0001-7285-373X, gname=O., sname=Ivanova]{O. Ivanova}
\affiliation{Astronomical Institute of the Slovak Academy of Sciences, 059 60 Tatransk\'{a} Lomnica, Slovak Republic}
\affiliation{Main Astronomical Observatory of the National Academy of Sciences of Ukraine, Kyiv, Ukraine}
\email{ivanova@email.com}

\author[orcid=0009-0000-1081-7944, gname=S., sname=Mykhailova]{S. Mykhailova}
\affiliation{Astronomical Observatory Institute, Faculty of Physics and Astronomy, Adam Mickiewicz University, Słoneczna 36, 60-286 Poznań, Poland}
\email{sofiia.mykhailova@gmail.com}

\author[orcid=0000-0001-5989-3630,gname=J., sname=Markkanen]{J. Markkanen}
\affiliation{Institut f\"ur Geophysik und Extraterrestrische Physik, TU Braunschweig, 38106 Braunschweig, Germany}
\email{markkanen@email.com}

\author[orcid=0000-0002-3779-7864,gname=I., sname=Lukyanyk]{I. Luk'yanyk}
\affiliation{Astronomical Observatory of Taras Shevchenko National University of Kyiv, 3 Observatorna St., Kyiv, 04053, Ukraine}
\email{lukyanyk@email.com}

\author[orcid=0000-0003-3462-5714,gname=T., sname=Kwiatkowski]{T. Kwiatkowski}
\affiliation{Astronomical Observatory Institute, Faculty of Physics and Astronomy, Adam Mickiewicz University, Słoneczna 36, 60-286 Poznań, Poland}
\email{kwiatkowski@email.com}

\author[orcid=0000-0002-5356-6433,gname=D., sname=Oszkiewicz]{D. Oszkiewicz}
\affiliation{Astronomical Observatory Institute, Faculty of Physics and Astronomy, Adam Mickiewicz University, Słoneczna 36, 60-286 Poznań, Poland}
\email{oszkiewicz@email.com}

\author[orcid=0000-0001-7403-1721,gname=A., sname=Penttila]{A. Penttil\"a}
\affiliation{Department of Physics, PO Box 64, FI-00014 University of Helsinki, Finland}
\email{penttila@email.com}

\author[orcid=0000-0002-9986-3898,gname=N., sname=Erasmus]{N. Erasmus}
\affiliation{South African Astronomical Observatory, 1 Observatory Rd, Observatory, Cape Town, 7925, South Africa}
\affiliation{Department of Physics, Stellenbosch University, Stellenbosch, 7602, South Africa}
\email{erasmus@email.com}

\author[orcid=0000-0002-7521-1078,gname=K., sname=Ergashev]{K. Ergashev}
\affiliation{Ulugh Beg Astronomical Institute, 33 Astronomicheskaya St., Tashkent, 100052, Uzbekistan}
\email{ergashev@email.com}

\author[orcid=0000-0002-3171-9873,gname=Y., sname=Krugly]{Y. Krugly}
\affiliation{LTE, Observatoire de Paris, Universit\'e PSL, Sorbonne Universit\'e, Universit\'e de Lille, LNE, CNRS, 75014, Paris, France}
\affiliation{V.N. Karazin Kharkiv National University, Sumska 35, Kharkiv, 61022, Ukraine}
\email{krugly@email.com}

\begin{abstract}
Interstellar objects (ISOs) provide direct constraints on the composition and evolution of extrasolar planetary systems. As the third confirmed ISO, comet 3I/ATLAS offers a rare opportunity to study pristine material from another star system. We present a systematic multi-band monitoring campaign conducted from 2025 July to September. To minimize systematic uncertainties, we utilized a homogeneous dataset of SDSS $g', r', i', z_s$ and Johnson-Cousins $BVR$ imaging obtained with MuSCAT3/4 on the 2.0-m Faulkes Telescopes and the 1.5-m Maidanak telescope. We employed a novel multi-aperture technique providing a robust, distance-independent characterization of the coma, alongside optical spectroscopy from SALT and the Nordic Optical Telescope. 3I/ATLAS exhibited a stable reddish color with no significant secular variation during its inbound journey. Spectroscopic analysis reveals carbon-chain depletion, with CN as the only prominent molecular emission. Coma morphology, analyzed via azimuthal averaging and rotational gradient filters, shows a persistent sunward jet at a position angle of $\sim$280$^{\circ}$, indicating sustained and localized outgassing from a high-latitude active region. Light-scattering modeling demonstrates that the observed colors and polarization are consistent with dense aggregates of organic matter and silicate minerals, while water ice sublimation remained insufficient to alter the optical properties. The stability of the coma structures and the homogeneous nature of our dataset indicate that 3I/ATLAS underwent a steady activation phase, preserving a volatile-rich but carbon-chain-poor composition. This characterization points to an object formed in a distinct protoplanetary environment and provides a definitive record of its pre-perihelion evolution.
\end{abstract}



\section{Introduction}

Interstellar objects (ISOs) offer a direct glimpse into the population of small bodies formed in planetary systems other than our own. Following 1I/`Oumuamua and 2I/Borisov, comet C/2025 N1 (3I/ATLAS, hereafter 3I) is the third confirmed ISO and the first discovered sufficiently early to permit extensive monitoring over a large fraction of its inbound trajectory \citep{Bolin25}. Its strongly hyperbolic orbit and clear cometary activity immediately established it as a prime target for coordinated observational campaigns.

Although 3I is observed as an unbound object rather than as an exocomet still orbiting its parent star, its study connects naturally with the broader exocomet literature. Exocomets are now recognized through several complementary channels: time-variable metal absorption during transits \citep{Kiefer2014}, asymmetric photometric dips produced by dusty tails crossing stellar disks \citep{Rappaport2018,Zieba2019}, and secondary gas in debris disks whose short-lived species require continuous replenishment from icy planetesimals \citep{Matra2017betaPic,Kral2017}. The archetypal case is $\beta$~Pictoris, where decades of Ca~II monitoring have revealed distinct dynamical families of star-grazing planetesimals \citep{Kiefer2014}, while ALMA measurements in this and other systems show CO+CO$_2$ ice fractions that can overlap those of Solar System comets \citep{Matra2017Fomalhaut}. Collectively, these observations demonstrate that comet-like bodies are common by-products of planet formation, but that their volatile content and dynamical architecture vary substantially among systems \citep{Strom2020,Iglesias2025}.

Interstellar comets provide a complementary route to the same problem because they allow the techniques of Solar System cometary astronomy to be applied directly to macroscopic material ejected from another planetary system. In contrast to transiting exocomets, whose nuclei and comae are unresolved and inferred through absorption or occultation signatures, 3I can be followed as an individual active body with spatially resolved coma morphology, broadband color measurements, gas production estimates, and dust-scattering constraints. Its pre-perihelion evolution therefore provides a bridge between the unresolved exocomet population and the detailed compositional taxonomy of Solar System comets.

Early observations showed that 3I was already active at large heliocentric distance, with a well-developed coma and pronounced dust structures \citep{deLaFuenteMarcos2025,2025ApJ...994L...3J,2025ATel17363....1B}. At the same time, the first optical spectra suggested that the comet was initially dominated by reflected dust continuum, with only weak or absent gas features in the visible \citep{deLaFuenteMarcos2025, Manzano2025}. Subsequent spectroscopic studies reported the emergence of CN emission and evidence for a composition atypical of many Solar System comets, while infrared observations indicated unusually high abundances of hypervolatile species, particularly CO$_2$, in the coma \citep{Schleicher2025,Cordiner2025}. These results suggest that 3I may preserve material formed under physical and chemical conditions distinct from those that prevailed in the protosolar nebula.

Within this broader context, the pre-perihelion evolution of the comet is of particular interest. Broadband colors constrain the optical properties of the dust, aperture photometry and the $Af\rho$ parameter trace the development of dust production, and morphological analysis can reveal persistent jets or fans associated with localized active regions on the nucleus \citep{Farnham2009}. When combined with spectroscopic information, these observables provide an important empirical basis for interpreting the activity of this interstellar comet.

An initial characterization of 3I was presented by \citet{Santana-Ros2025}, who reported its rotation period, spectra, and preliminary color measurements. In this work, we extend that analysis by providing a systematic multi-band photometric monitoring campaign and a detailed investigation of the coma morphology covering the steady pre-perihelion evolution. By leveraging standardized MuSCAT-class instrumentation and consistent filter sets across the Faulkes Telescopes and dedicated monitoring at Maidanak Observatory, we provide a homogeneous dataset that minimizes instrumental systematics over a significantly longer temporal baseline. For this analysis, we adopt a new photometric approach that accounts for the specific observing conditions of each image dataset while retaining standardized measurements within a fixed projected cometocentric radius of 10,000 km. Furthermore, we incorporate new optical spectroscopy obtained with the Southern African Large Telescope (SALT) and the Nordic Optical Telescope (NOT). This spectral analysis seeks to identify the onset of emission features, particularly CN, at smaller heliocentric distances where such activity is expected to become detectable following its non-detection in our earlier data obtained at larger distances. Our objective is to characterize the comet's color and brightness evolution and examine the stability of inner-coma structures to better constrain the origin and physical nature of this interstellar visitor. To constrain physical properties of dust particles, we model light-scattering features of dust in the coma using an aggregated particle model with varying composition and structure. 

This paper is organized as follows. In Sect.~\ref{sec:observations}, we describe the multi-band photometric and spectroscopic observations obtained with the Faulkes Telescopes, Maidanak Observatory, SALT, and the NOT, along with the data reduction and calibration procedures. In Sect.~\ref{sec:results}, we present the results of our monitoring campaign, including the evolution of the comet's brightness and color, its gas production, and a morphological analysis of the inner coma. Within this section, we also describe the light-scattering models used to constrain the physical properties of the dust. Finally, our main findings are summarized and discussed in Sect.~\ref{sec:conclusions}.

\section{Observations and Data Reduction}
\label{sec:observations}

The observations presented here constitute a temporal extension of the monitoring campaign described in \citet{Santana-Ros2025}, now encompassing the comet’s entire steady pre-perihelion approach. While leveraging the same core observational facilities—the Maidanak Observatory in Uzbekistan and the Las Cumbres Observatory nodes at Haleakal\={a} (FTN) and Siding Spring (FTS)—the present study implements a refined photometric processing strategy. This updated approach allows for a more robust characterization of the coma evolution and provides a more homogeneous baseline for interpreting activity levels than the initial post-discovery analysis. 

The combined campaign spans a two-month interval of the comet's inbound trajectory, from late July to mid-September 2025. Photometric monitoring with the Faulkes 2.0-m Telescopes began on 2025 July 26 and continued through 2025 September 11. This was complemented by a high-cadence campaign at the Maidanak Observatory from 2025 August 1 to 29, providing intensive $BVR$ coverage during a period of rapidly increasing activity. The detailed logs of observations, including heliocentric ($r_h$) and geocentric ($\Delta$) distances, are provided in Tables~\ref{tab:maidanak_obs} and \ref{tab:lco_obs}.

\subsection{Photometry}
A primary asset of this campaign is the instrumental homogeneity of the Faulkes Telescopes data. We utilized the 2.0-m Faulkes Telescope North (FTN; MPC code F65) and the 2.0-m Faulkes Telescope South (FTS; MPC code E10) to obtain photometry in the SDSS $g', r', i'$, and $z_s$ bands using the MuSCAT3 and MuSCAT4 instruments, respectively \citep{Narita2020, Fukui2022}. Both instruments enable simultaneous multi-band imaging via dichroic beam splitters, allowing us to obtain a consistent photometric dataset from both hemispheres without the need for inter-instrument color-term corrections. This approach significantly reduces the systematic noise floor in our analysis of color evolution.

MuSCAT3 is equipped with four 1k~$\times$~1k CCDs (Hamamatsu Photonics), each with a pixel scale of 0.27\arcsec\ and a field of view (FOV) of 7.4~$\times$~7.4~arcmin$^2$. MuSCAT4 features four 2k~$\times$~2k CCDs with a pixel scale of 0.38\arcsec\ and an FOV of 13~$\times$~13~arcmin$^2$. The detectors exhibit low readout noise ($\sim$3--5~e$^-$). The photometric data were processed using the MuSCAT pipeline, which includes standard calibrations (bias, dark, and flat-field corrections) and aperture photometry optimized for each band.

Complementary Johnson-Cousins $B, V, R$ imaging was conducted in August 2025 at the 1.5\,m Ritchey-Chr\'{e}tien (RC) telescope (AZT-22) at the Maidanak Observatory (hereafter Maidanak), located at an altitude of approximately 2650\,m in southeastern Uzbekistan \citep{2018NatAs...2..349E}. This facility provided high-cadence monitoring that overlapped with the Faulkes Telescopes campaign, allowing for a broader spectral characterization of the coma. Imaging sequences were acquired with the SNUCAM instrument, featuring a Spectral Instruments 600 back-illuminated $4096\times4096$ pixel CCD camera providing a field of view of $18\farcm1\times18\farcm1$ and delivering a pixel scale of $0\farcs26$\,pixel$^{-1}$ \citep{MaidanakCCD}. Standard reduction procedures---including bias and flat-field correction---were applied to the Maidanak data, following the same methodology used for the Faulkes Telescopes dataset.

\subsubsection{Image reduction and photometric analysis}

\begin{figure*}
\centering
\includegraphics[width=0.48\columnwidth]{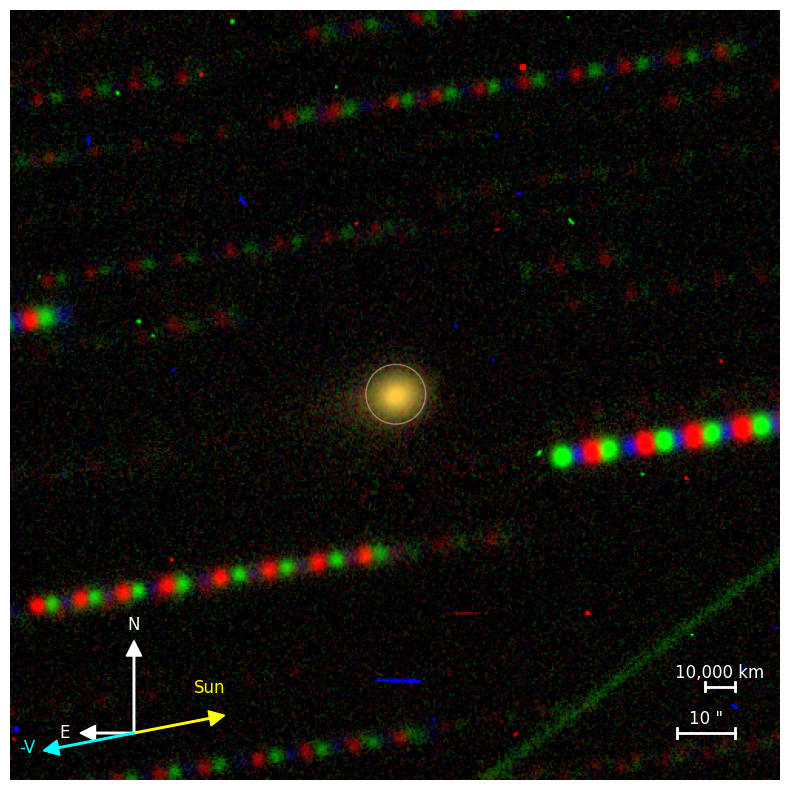}
\includegraphics[width=0.48\columnwidth]{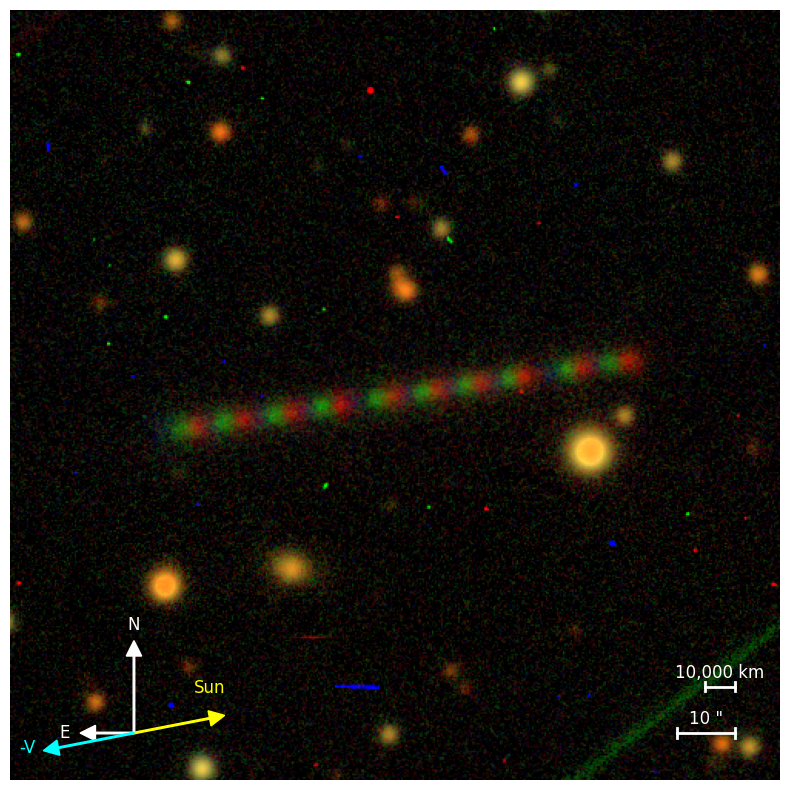}
\caption{Example of the stacking procedure for the Maidanak data obtained on 2025 August 15, based on ten 60~s exposures in each of the $B$, $V$, and $R$ filters. \textbf{Left:} Comet-centered stack, showing the coma morphology without trailing of the comet signal. The pink circle indicates an aperture radius of 10\,000~km. \textbf{Right:} Sidereal stack used for photometric calibration, in which the comet appears trailed.}
\label{fig:stacking_example}
\end{figure*}

To maximize the signal-to-noise ratio of the comet measurements, we developed a dedicated pipeline in which the individual exposures were combined into deep stacks in each filter using two complementary strategies (Fig.~\ref{fig:stacking_example}).

For the comet analysis, consecutive exposures were aligned in the comet reference frame using topocentric ephemerides from the JPL Horizons system, evaluated at the mid-time of each exposure. These ephemerides provided the comet position, geocentric distance, $\Delta$, and apparent sky motion.

The individual exposures were reprojected onto a common grid centered on the comet using exact, flux-conserving WCS transformations. Before combining these exposures, we removed resolved field stars. This step was particularly crucial during the early epochs when 3I/ATLAS was projected against relatively crowded stellar fields. Because the telescope tracked sidereally, the stars remained point-like in the individual frames while the comet moved between exposures. In each reprojected image, we simultaneously fit a constant background, a 2D Gaussian for the comet, and 2D Moffat profiles for all stars from the Gaia DR3 catalog within a $0.5\,\text{arcmin}$ radius. Only the fitted Moffat components were subtracted. Finally, the cleaned exposures were combined into comet-centered stacks, preserving the cometary signal while preventing stars at varying cometocentric positions from contaminating the photometric apertures.

The same exposures were also combined in a fixed celestial reference frame to produce sidereal stacks. In these images, the background stars remained point-like and could therefore be measured reliably for photometric calibration, whereas the comet appeared trailed.

Photometric calibration was derived from the sidereal stacks using field stars matched to the Pan-STARRS catalog \citep{2020ApJS..251....6M}. After centroid refinement, stellar fluxes were measured with circular apertures, and the local sky background was estimated from surrounding annuli using sigma-clipped statistics. To enable a consistent comparison of comet fluxes obtained under different seeing conditions, stellar growth curves were constructed from isolated reference stars and used to determine the encircled-energy fraction as a function of aperture radius. These corrections allowed comet photometry measured in apertures comparable to the image full width at half maximum (FWHM) to be placed on a homogeneous relative scale between filters.

Comet photometry was performed on the comet-centered stacks. The photocentre was refined around the nominal ephemeris position by centroiding within a seeing-scaled region; if the derived shift was unphysically large, the nominal center was adopted instead. Circular-aperture photometry was then measured in a series of apertures corresponding to fixed projected cometocentric radii. These were converted from kilometers to angular and pixel units using the contemporaneous geocentric distance from the ephemerides. The local coma background was estimated on a surrounding annulus using sigma-clipped statistics. Calibrated comet magnitudes were then obtained from the background-subtracted fluxes using the stellar zero-points. This approach ensured that the stellar calibration and comet measurements were performed on the image products best suited to each task: sidereal stacks for the stars and comet-centered stacks for the moving target. The corresponding apparent magnitudes measured within the same $10\,000$~km aperture, together with their $1\sigma$ uncertainties, are listed in Tables~\ref{tab:lco_photometry} and \ref{tab:maidanak_photometry} for the Faulkes Telescopes and Maidanak datasets, respectively.

\begin{figure}
    \centering
    \includegraphics[width=0.58\linewidth]{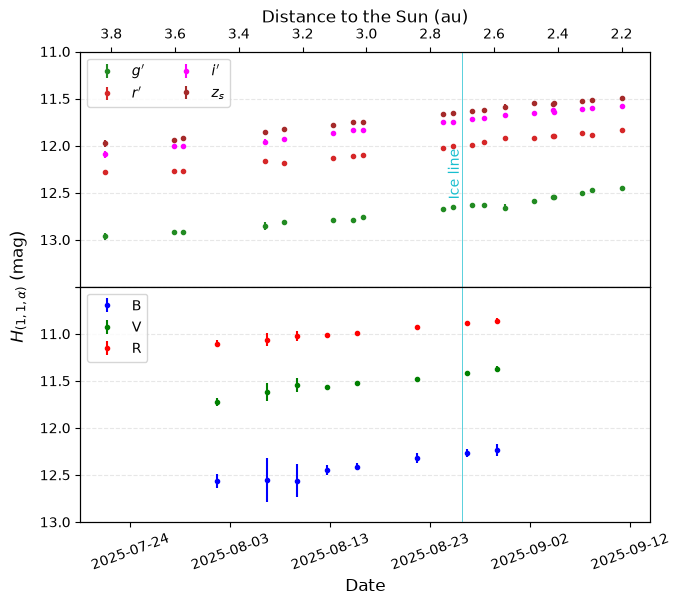}
   \caption{Multiband $H(1,1,\alpha)$ magnitude evolution of comet 3I/ATLAS during its pre-perihelion approach. The top x-axis indicates the heliocentric distance ($r_h$) in au, while the bottom x-axis shows the observation dates. \textbf{Top panel:} Photometry in the SDSS $g', r', i',$ and $z_s$ filters obtained with the Faulkes 2.0-m telescopes. \textbf{Bottom panel:} Johnson-Cousins $B, V,$ and $R$ photometry acquired at the Maidanak 1.5-m telescope. Error bars denote 1$\sigma$ photometric uncertainties, which are dominated by the atmospheric conditions at the time of the observations. Both datasets show a consistent brightening trend as the comet approaches perihelion, with the coverage extending from $r_h \approx 3.7$~au to $r_h \approx 2.1$~au}.
    \label{fig:photometry}
\end{figure}

Figure~\ref{fig:photometry} presents the corresponding distance-normalized photometric results for 3I/ATLAS.
The plotted values correspond to absolute magnitudes $H(1,1,\alpha)$, corrected for the heliocentric and geocentric distances, thereby removing the dominant effect of the changing observing geometry.
We retrieved ephemerides using the JPL Horizons system \citep[see \url{https://ssd.jpl.nasa.gov/horizons/}]{Giorgini1996} via the astroquery API \citep{Ginsburg2019}. The upper panel illustrates the brightness evolution in the SDSS $g', r', i',$ and $z_s$ filters using data from the Faulkes Telescopes, while the lower panel shows the Johnson-Cousins $B, V,$ and $R$ magnitudes obtained at the Maidanak Observatory. The data are plotted as a function of both time (bottom axis) and heliocentric distance $r_h$ (top axis), providing a consistent view of the comet's steady brightening across the observed inbound trajectory.

Stacked images of comet 3I in the SDSS $r'$ filter were carefully processed using digital enhancement techniques to reveal morphological structures within the coma. To enhance faint features, we applied digital filters, including the rotational gradient method developed by \citet{larson1984coma} and division by the azimuthal average \citep{samarasinha2014image}, allowing for the detection of subtle morphological details. The image processing and digital enhancement were performed using our in-house IDL scripts. Because different enhancement techniques modify images in distinct ways, we applied each digital filter both to individual frames and to the combined image. This approach allowed us to verify whether the detected features were consistent and therefore likely to be real, rather than artifacts introduced by the processing. In this way, we minimize the risk of misinterpreting spurious structures produced by the filtering methods as genuine. We also examined the stability of the jet morphology with respect to shifts in the comet’s optocenter. Even with noticeable displacements, the jet structure remained unchanged, supporting its physical origin. The results of this analysis are illustrated in Fig.~\ref{fig:morphol}.

\subsection{Spectroscopy}
\label{sect:spectroscopy}

Spectroscopic observations were carried out using the same facilities as described in \citet{Santana-Ros2025}, the 10-m Southern African Large Telescope (SALT) and the 2.56-m Nordic Optical Telescope (NOT). 

SALT spectra were acquired on 2025 August 29 with the Robert Stobie Spectrograph (RSS) under the program 2025-1-DDT-004 (PI: T. Kwiatkowski). A total of 15 exposures, each with an integration time of 120~s, were obtained using the PG0700 grating and a $0.6''$ slit aligned along the cometary trail with the atmospheric dispersion corrector in use. The observations were performed in $2 \times 2$ binning mode, resulting in a spectral resolving power of $R \sim 1400$. The final combined spectrum covers the wavelength range 3500--7400~\AA, with gaps at 4820--4980~\AA\ and 6160--6300~\AA. During the same night, a spectrum of the solar analog star BD-004074 (SA~112-1333, \citealt{Landolt1973}) was obtained at a comparable airmass and was used for the removal of the dust-reflected solar continuum.

Additional spectra were obtained with the ALFOSC instrument mounted on the NOT on 29 July, 29 August, and 3 September 2025 under program 71-411 (PI: A. Penttil\"a) together with spectra of the solar analog star BD-004074 \citep{landolt1992ubvri}. Different grisms were used for each epoch (\#4, \#20, and \#6, respectively). Owing to a technical issue, an incorrect grism was employed on August 29; nevertheless, the resulting spectrum, covering the wavelength range 5650--10150~\AA, was retained and used to estimate the spectral slope in the optical wavelength region.

Data reduction and calibration were performed using standard procedures within the IRAF packages \citep{IRAF}, including bias subtraction, flat-field correction, and wavelength calibration using arc lamp exposures. Individual spectra of 3I were median-combined and subsequently divided by the median-combined spectrum of the solar analog star. The NOT spectra were additionally flux-calibrated using observations of the spectrophotometric standard star Feige~110 \citep{oke1990faint}
obtained during the same nights.

A summary of the spectroscopic observations and the corresponding instrumental parameters is provided in Table~\ref{tab:spectra_obs}.

\section{Results}
\label{sec:results}

\subsection{Color evolution and radial gradients}

\begin{figure}
    \centering
    \includegraphics[width=0.58\linewidth]{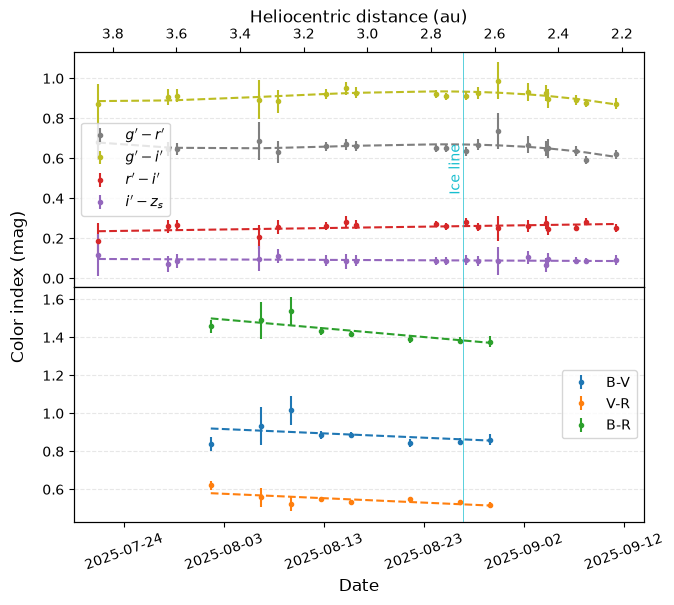}
    \caption{Color evolution of comet 3I/ATLAS during the pre-perihelion phase. \textbf{Top:} SDSS color indices $(g' - r')$, $(g' - i')$, $(r' - i')$, and $(i' - z_s)$ derived from the Faulkes Telescopes monitoring. \textbf{Bottom:} Johnson-Cousins $(B - V)$, $(V - R)$ and $(B - R)$ color indices obtained from the Maidanak dataset. The top x-axis represents the heliocentric distance $r_h$. The dashed vertical line indicates the water-ice snow line (at $r_h \approx 2.7$~au), marking the region where water-ice sublimation typically begins to dominate the activity. Error bars represent 1$\sigma$ uncertainties propagated from the photometric measurements.} 
    \label{fig:color_evolution}
\end{figure}

We analyzed the temporal evolution of the optical colors of 3I measured within a projected aperture radius of 10,000 km from the comet center (Fig.~\ref{fig:color_evolution}). Over the monitored pre-perihelion interval, the color indices remained nearly constant, with no evidence of strong secular variations as the comet approached the Sun. The $(g'-i')$ color is systematically the reddest, staying close to $\sim$0.9 mag, and showing a similar evolution to $(g'-r')$, while $(r'-i')$ remains near $\sim$0.25 mag and $(i'-z_s)$ near $\sim$0.09 mag.

The slight change in $g'-i'$ and $g'-r'$ after crossing the adopted ice-line distance may suggest that the shorter-wavelength coma reflectance became more sensitive to the increasing activity level. This behavior could reflect a change in the dust population, such as a larger contribution from finer or fresher grains released as water-ice sublimation became more efficient, while the redder indices remained comparatively stable. A contribution from weak gas emission affecting the $g'$ band cannot be excluded. In our spectra, CN emission lines are detected (see Sect. \ref{sec:spectra_analysis} for details), indicating the presence of gaseous activity. However, our estimates indicate that the emission contribution to the photometric measurements is small: as an order-of-magnitude estimate for an aperture radius of $\sim$10\,000 km, it is $\sim2.1\%$ in the SDSS $g'$ filter and $\sim1.3\%$ in the $r'$ filter. The gas emission contribution was estimated by convolving the observed spectrum and the corresponding solar analog spectrum as a continuum with the SDSS filter transmission curves, using the data on September 3 as a representative case.

\begin{figure}
    \centering
    \includegraphics[width=0.58\linewidth]{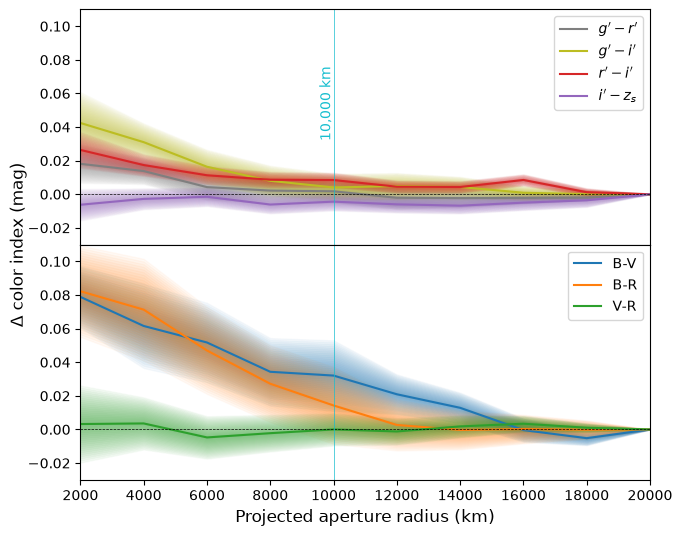}
    \caption{Radial color profiles of comet 3I/ATLAS. The plot shows the median color indices $(g'-r')$, $(g'-i')$, $(r'-i')$, and $(i'-z_s)$ (top panel), and $B - V$, $B - R$, $V - R$ (bottom panel) as a function of projected aperture radius, normalized to their values at $20\,000$ km. Shaded regions indicate the uncertainties of the median color estimates. The vertical line marks the reference aperture of $10\,000$ km used in the photometry analysis.
    }
\label{fig:color_radius}
\end{figure}

Figure~\ref{fig:color_radius} shows the median radial color variation of 3I relative to the value measured at $20\,000$~km. Both $(g' - i')$ and $(r' - i')$ are positive at small cometocentric distances and gradually decrease towards zero with increasing radius. This indicates that the inner coma is redder than the outer coma in these color indices. In contrast, $(i' - z_s)$ remains close to zero over the full radial range, showing no significant gradient within the uncertainties. The strongest relative color gradient is thus observed at shorter wavelengths, while the reddest part of the optical range appears nearly uniform.

The published color measurements show that 3I was overall red in the optical, but with significant scatter between epochs and datasets. In the earliest near-discovery data, \citet{Seligman25} reported SDSS-like color indices (see Table~\ref{tab:color_comparison}), while their precovery ZTF photometry gave much bluer values in May and June 2025. 

\citet{Bolin25} found a range of Johnson-Cousins and SDSS colors, while \citet{Kareta2025} measured the $g^\prime-i^\prime$ index. On 15 July 2025, \citet{Beniyama2025} measured several color indices, whereas the month-long monitoring of \citet{Santana-Ros2025} yielded specific campaign-averaged colors, as shown in Table~\ref{tab:color_comparison}.

In this work, we obtain average colors within a projected aperture of $10\,000$~km for both the Johnson-Cousins ($B, V, R$) and the broadband SDSS $g', r', i', z_s$ filters. The mean values obtained for the complete observation period are $(g' - r') = 0.644 \pm 0.006$, $(r' - i') = 0.265 \pm 0.005$, $(i' - z_s) = 0.089 \pm 0.006$ and $(g' - i') = 0.909 \pm 0.007$. These values are redder than solar and are in good agreement with the intermediate July measurements, especially the \citet{Santana-Ros2025} campaign averages and the \citet{Beniyama2025} spectrophotometry, while remaining somewhat less red than the earliest near-discovery colors reported by \citet{Seligman25} and \citet{Bolin25}.

The significant scatter between different datasets may be attributed to a combination of factors, including differences in aperture size, instrumental systematics, and the intrinsic variability of the coma as it evolved during the pre-perihelion approach. Our homogeneous dataset, obtained with consistent instrumentation and reduction procedures, provides a stable baseline for interpreting the color evolution, suggesting that while the comet's optical color was generally red, it did not exhibit strong secular changes over the monitored period.

\begin{table*}
\centering
\footnotesize 
\setlength{\tabcolsep}{2.2pt} 
\caption{Comparison of observed color indices for comet 3I/ATLAS.}
\label{tab:color_comparison}

\begin{tabular}{l c c c c c c c c}
\hline\hline
Reference & Date & $g'-r'$ & $r'-i'$ & $i'-z_s$ & $g'-i'$ & $B-V$ & $V-R$ & $R-I$ \\
 & (2025) & & & & & & & \\
\hline
\citetalias{Seligman25} & May 22\tablenotemark{a} & $0.42\pm0.14$ & -- & -- & -- & -- & -- & -- \\
\citetalias{Seligman25} & Jun 18\tablenotemark{a} & $0.44\pm0.14$ & -- & -- & -- & -- & -- & -- \\
\citetalias{Seligman25} & Post-d. & $0.85\pm0.03$ & $0.25\pm0.03$ & $0.20\pm0.08$ & -- & -- & -- & -- \\
\citetalias{Bolin25}    & July    & $0.84\pm0.05$ & $0.16\pm0.03$ & $-0.02\pm0.07$ & $1.00\pm0.05$ & $0.98\pm0.23$ & $0.71\pm0.09$ & $0.14\pm0.10$ \\
\citetalias{Kareta2025} & July    & --             & --             & --              & $0.98\pm0.03$ & -- & -- & -- \\
\citetalias{Beniyama2025}& Jul 15  & $0.603\pm0.031$ & $0.210\pm0.031$ & $0.117\pm0.046$ & -- & -- & -- & -- \\
\citetalias{Santana-Ros2025}& July & $0.65\pm0.03$ & $0.27\pm0.03$ & $0.10\pm0.04$ & $0.91\pm0.03$ & -- & $0.42\pm0.02$ & -- \\
\hline
\textbf{This work\tablenotemark{b}} & \textbf{Jul--Sep} & $\mathbf{0.644\pm0.006}$ & $\mathbf{0.265\pm0.005}$ & $\mathbf{0.089\pm0.006}$ & $\mathbf{0.909\pm0.007}$ & $\mathbf{0.867\pm0.009}$ & $\mathbf{0.542\pm0.005}$ & -- \\
\hline
\end{tabular}

\tablenotetext{a}{Precovery ZTF photometry.}
\tablenotetext{b}{Mean values calculated over the complete observation period.}
\end{table*}

\subsection{Coma morphology and jet stability}
\begin{figure}
\centering
\includegraphics[angle=0,width=0.48\columnwidth]{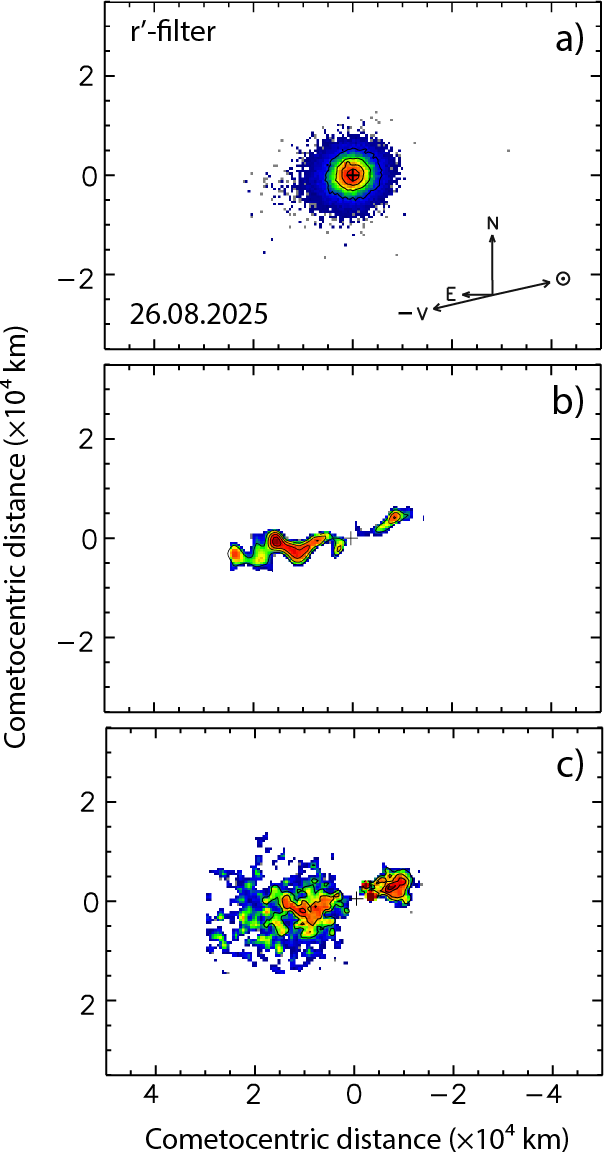}   
\caption{Morphological features revealed in the coma of comet 3I/ATLAS. Different images contain the results of processing by various digital filters: (a) the summed image of the comet in the SDSS $r'$ filter; (b) the Larson-Sekanina filter; (c) division by the azimuthal average. The black cross denotes the optocenter of the comet. Black arrows show directions of North (N), East (E), sunward ($\odot$), and the comet's negative velocity vector.
\label{fig:morphol}}
\end{figure}

The analysis of morphological structures and their temporal evolution provides important insights into nucleus rotation, activity variations, physical properties, and the interactions between gas and dust \citep{Farnham2009}. Our analysis of the morphology of 3I shows (see Fig.~\ref{fig:morphol}) that after applying image enhancement techniques, a large-scale structure is revealed, dominated by a relatively stable sunward jet at a position angle measured east of north of about $278^\circ\pm2^\circ$, while the dust tail in the anti-solar direction appears at $106^\circ\pm4^\circ$.

Moreover, pre-perihelion observations in August 2025 by \citet{2026A&A...705L...3S} applied Laplacian filtering to high-cadence imaging obtained with the Two-meter Twin Telescope and independently confirmed the presence of a narrow high-latitude jet in the inner coma. High-precision measurements of the jet position angle at a projected distance of $6000\,km$ from the optocenter yield a mean value of $280.7 \pm 0.2^\circ$, fully consistent with our data. These authors also detected a small periodic modulation of the jet PA with a period of $7.74 \pm 0.35$\,h, which they interpret as the precessional motion of a high-latitude jet around the sky-projected spin axis of the nucleus.

The persistence of both the sunward jet and the anti-solar tail \citep{2025ATel17372....1I} is corroborated by independent observations with the 1.0-m Lesedi telescope at the South African Astronomical Observatory (SAAO, South Africa). Comparison with images reported by \citet{2025ATel17363....1B} further confirmed the sunward structure, indicating sustained localized outgassing on the nucleus. These values are also consistent with jet orientations reported by other observers: early imaging on 2 July with the 2-m telescope revealed a plume at a position angle near $279.5^\circ$ \citep{2026A&A...705L...3S}, and subsequent Hubble Space Telescope observations confirmed a similar morphology, showing a well-resolved fan-like feature at $280^\circ$ \citep{2025ApJ...994L...3J}.

Overall, the data indicate that the sunward jet is a relatively stable and persistent structure, observable over several months, from early July through September, while the anti-solar tail remains clearly defined under the influence of solar radiation. This stability is now understood to arise from a discrete high-latitude active region (latitude $\approx +75^\circ$), whose emission cone projects as the observed sunward fan/plume \citep{2025ATel17350....1O}.

\subsection{Spectral slopes and emission features}
\label{sec:spectra_analysis}
Spectral slopes were derived from reflectance spectra obtained by dividing the observed spectra of 3I by those of the solar analog star, BD-004074, observed during the same night. The resulting reflectance spectra were normalized to unity at 5500~\AA, and spectral gradients were measured by performing linear fits over selected wavelength intervals. This methodology ensures consistency with our previous study of 3I \citep{Santana-Ros2025} based on observations from July 2025, which showed a strong reddening in the visible range. The new August–September observations reveal noticeably less red (flatter) slopes over the same wavelength intervals. The derived slopes, including new measurements obtained between 15 July and 3 September 2025, are presented in Table~\ref{tab:sp_slopes2} and provide a consistent record of the continuum evolution as the object approached the water-ice sublimation line.

To search for molecular emissions, we compared the observed SALT spectrum obtained on August 29 (Fig.~\ref{fig:fig_sp_September}) with calculated spectra of typical cometary species \citep{Kim1994}. For each spectral feature, we estimated the local continuum mean and its standard deviation ($\sigma$) over a window of $\pm 20$~\AA, excluding the line region. The line signal was defined as the continuum-subtracted flux. The signal-to-noise ratio (S/N) was computed as the ratio of the peak line signal to $\sigma$. In addition, we calculated the integrated signal-to-noise ratio:
\begin{equation}
\mathrm{S/N}_{\mathrm{int}} = \frac{F_{\mathrm{line}}}{\sigma \sqrt{N}},
\end{equation}
where $F_{\mathrm{line}}$ is the integrated line flux and $N$ is the number of spectral bins across the feature. We adopted a detection threshold of $\mathrm{S/N} \geq 3$ for robust detections.

Applying this method, the CN ($\mathrm{B}^2\Sigma^+ - \mathrm{X}^2\Sigma^+$, $\Delta \nu = 0$) band at $\sim 3883.5$~\AA\ reaches $\mathrm{S/N} \approx 4.5$ and $\mathrm{S/N}_{\mathrm{int}} \approx 5.14$, constituting a secure detection. The C$_3$ emission shows a marginal-to-moderate detection with peak $\mathrm{S/N} \approx 3.2$ and $\mathrm{S/N}_{\mathrm{int}} \approx 2.97$. The CO$^+$ ($\mathrm{A}^2\Pi - \mathrm{X}^2\Sigma^+$ (2,0)) band at 4260--4290~\AA\ is not significant, with peak $\mathrm{S/N} \approx 1.5$ and $\mathrm{S/N}_{\mathrm{int}} \approx 0.50$. Similarly, the N$_2^+$ ($\mathrm{B}^2\Pi - \mathrm{X}^1\Sigma^+$, $\Delta \nu = 0$) feature at 3914~\AA\ yields peak $\mathrm{S/N} \approx 2.6$ and $\mathrm{S/N}_{\mathrm{int}} \approx 1.13$, remaining below the detection threshold.

The identification of C$_3$ remains uncertain, as its signal-to-noise ratio is near the detection threshold. Furthermore, C$_3$ is typically observed at smaller heliocentric distances than CN (which can be detected beyond 3~au) and C$_2$ (which generally appears near 2.5~au). Given that our observations were obtained at a heliocentric distance of 2.6~au, the absence of C$_2$ emission suggests that the marginal C$_3$ signal is likely spurious.

\begin{table}[ht!]
    \centering
    \caption{Evolution of the pre-perihelion spectral slope derived from SALT and NOT observations}
    \label{tab:sp_slopes2}
    
\centering
\begin{tabular}{c c c c c}
\hline\hline
Date & $r$ & Wavelength & Slope & Telescope \\
(2025) & (au) & (\AA) & (\%/1000\,\AA) & \\
\hline
Jul 15\tablenotemark{*} & 4.05 & 3600--5000 & $15.49\pm0.15$ & SALT \\
            & 4.05 & 5000--7200 & $21.19\pm0.04$ & SALT \\
Jul 25\tablenotemark{*} & 3.72 & 4000--5000 & $18.18\pm0.08$ & NOT \\
            & 3.72 & 5000--7000 & $22.89\pm0.17$ & NOT \\
            & 3.72 & 7000--9000 & $9.78\pm0.35$ & NOT \\
Jul 29      & 3.59 & 4000--5000 & $30.9\pm1.0$ & NOT \\
            & 3.59 & 5000--7000 & $25.8\pm0.2$ & NOT \\
            & 3.59 & 7000--9000 & $10.7\pm0.4$ & NOT \\
Aug 29      & 2.59 & 3600--5000 & $16.3\pm0.6$ & SALT \\
            & 2.59 & 5000--7200 & $19.5\pm0.5$ & SALT \\
            & 2.59 & 7000--9000 & $10.6\pm0.4$ & NOT \\
Sep 03      & 2.45 & 3600--5000 & $22.3\pm0.4$ & NOT \\
\hline
\end{tabular}
\tablenotetext{*}{Data from \citet{Santana-Ros2025}.}
    
    
\end{table}

\subsection{Estimation of CN production rate}
\begin{figure}
\centering
\includegraphics[width=0.58\columnwidth]{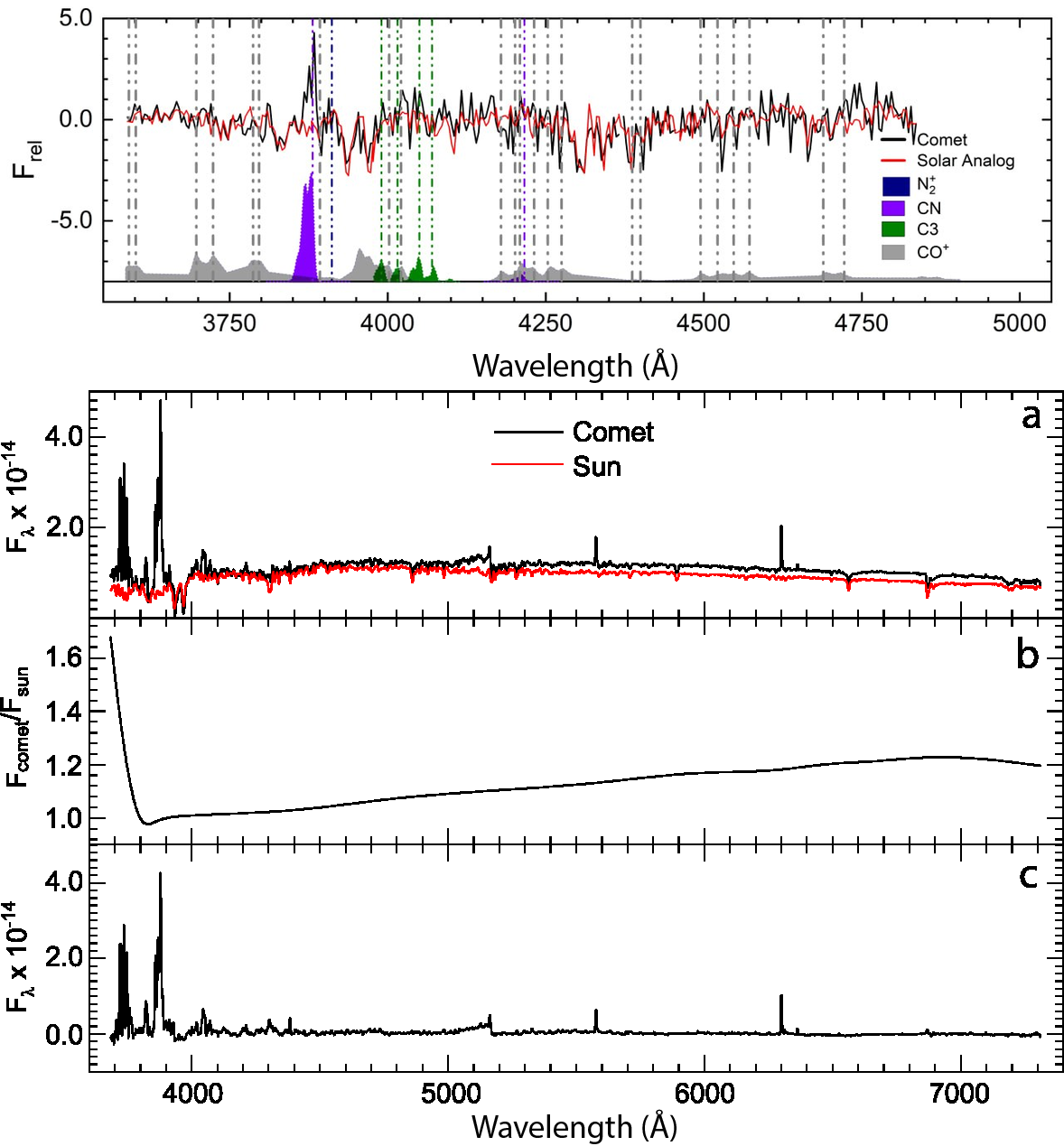}
\caption{Optical spectra of the 3I/ATLAS comet obtained on 29 August and 3 September 2025 with the SALT and NOT telescopes, respectively. \textbf{Top}: the SALT data. The black line shows the comet spectrum, and the red dashed line represents the solar-analog reflectance spectrum (both in relative units). In the lower part of this section, the calculated spectra of cometary emissions CN, C$_3$, CO$^+$, and N$_2^+$ are given. \textbf{Bottom}: the NOT data. Sub-panel (a) shows the spectrum of the comet (black line) with the scaled solar-analog spectrum (red line); (b) is the normalized spectral dependence of the dust reflectivity; (c) shows emission spectrum of the comet. The fluxes $F_{\lambda}$ are given in units of $\mathrm{erg\,s^{-1}\,cm^{-2}\,\mbox{\AA}^{-1}}$. }
\label{fig:fig_sp_September}
\end{figure}

Figure~\ref{fig:fig_sp_September} illustrates the procedure for deriving the pure emission spectrum. The reduced spectrum of comet 3I displays a number of weak structures that may be associated with the CO$^+$ ($A^2\Pi - X^2\Sigma^+$) system, though these remain near the detection limit.

Gas production rates for CN, along with $3\sigma$ upper limits for C$_2$ and C$_3$, were computed using a \citet{haser1957} model. Following standard methodology, we adopted fluorescence efficiencies ($g$-factors) and parent/daughter scale lengths from \citet{2011Icar..213..280L}. The CN $g$-factor, which depends on the Swings effect, was computed using the tables of \citet{2010AJ....140..973S} for the observed heliocentric distance $r = 2.444$~au and heliocentric velocity $v_h = -51.47$~km~s$^{-1}$. We derived a logarithmic gas production rate for CN of $\log Q(\text{CN}) = 25.6 \pm 0.2$ [mol s$^{-1}$], with $3\sigma$ upper limits of 22.6 and 24.3 for C$_3$ and C$_2$, respectively. Based on the calculated upper limits of the production rates, the logarithmic ratios are $\log(Q(\text{C}_2)/Q(\text{CN})) < -1.3$ and $\log(Q(\text{C}_3)/Q(\text{CN})) < -3.0$. These values fall significantly below the range of typical comets defined by \citet{Ahearn1995}, placing the comet in the "carbon-chain depleted" category according to the \citeauthor{Ahearn1995} and \citeauthor{2011Icar..213..280L} taxonomic classification.

These results align with contemporary reports: CN was reliably detected in late August 2025, with reported logarithmic production rates of $24.95 \pm 0.25$ \citep{Schleicher2025} and $23.9$ \citep{Manzano2025}. This chemical signature is reminiscent of the Jupiter-family comet 43P/Wolf–Harrington, where CN is often the only prominent molecular emission \citep{Schleicher1993}. Consequently, 3I can be classified as a "carbon-chain depleted" comet. This depletion is particularly interesting in light of JWST observations \citep{Cordiner2025}, which reported a high CO$_2$/H$_2$O ratio. The combination of carbon-chain depletion in the optical and CO$_2$ enrichment in the infrared suggests a volatile composition that differs significantly from typical Solar System comets, potentially reflecting a formation environment enriched in CO$_2$ ice.

\subsection{Light scattering model}

We model light scattering by various types of dust particles of different compositions. The dust particles are modeled as ballistic cluster-cluster (BCCA), particle-cluster (BPCA), and particle-cluster with two migrations (BAM2) aggregates \citep{Shen2008}. The aggregates were generated with our in-house Python script. The monomer radius is $r = 0.13$\,µm, and the number of monomers varies from 1024 to 65536. Five different compositions are tested: Mg$_{0.95}$-Fe$_{0.05}$SiO$_3$ pyroxene \citep{Dorschner1995}, Fe-rich olivine, weakly and heavily processed organic matter \citep{jenniskens1993}, and water ice \citep{warren1984}. The light-scattering characteristics for each aggregate, averaged over 10 random realizations, are computed using the fast superposition T-matrix method (FaSTMM2) \citep{Markkanen2017, Markkanen2026}. Three different wavelengths are considered $\lambda = 445, 550, 660$\,nm corresponding to $B-$, $V-$, and $R-$ filters, respectively.      

\begin{figure}
    \centering
    \includegraphics[width=0.58\linewidth]{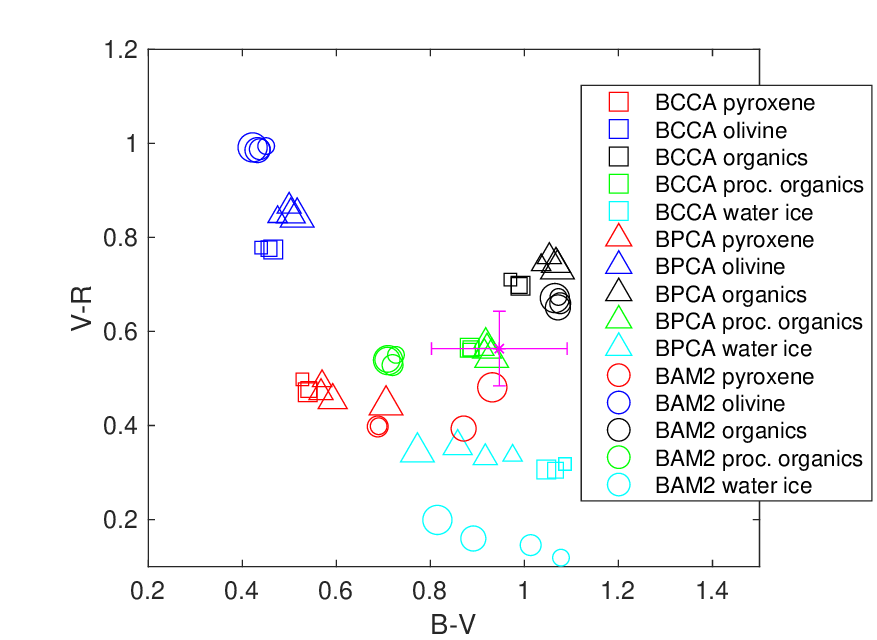}
    \caption{Modeled $(B-V)$ and $(V-R)$ colors for different types of aggregates. The magenta asterisk (\textasteriskcentered) indicates the mean $(B-V)$ and $(V-R)$ colors observed in this work, with error bars representing the $1\sigma$ standard deviation. The modeled colors (open symbols) represent different aggregate structures: Ballistic Cluster-Cluster Aggregates (BCCA), Ballistic Particle-Cluster Aggregates (BPCA), and BAM2 models. Different materials are indicated by color: pyroxene (red), olivine (blue), organics (black), processed organics (green), and water ice (cyan).}
    
\label{fig:color_model}
\end{figure}

Figure \ref{fig:color_model} shows the modeled $(B-V)$ and $(V-R)$ colors averaged over the phase angle range of 0--20 degrees for different aggregates. Marker sizes correspond to aggregate sizes; the smallest markers indicate aggregates of 1024 monomers, whereas the largest markers correspond to 65536 monomers.   

As seen in Fig. \ref{fig:color_model}, the aggregates made of processed organic matter show colors consistent with the observations. However, these aggregates do not reproduce polarimetric pre-perihelion observations reported by \citet{Gray2025}. Instead, a linear mixture of BAM2 aggregates made of organics (90\% in volume) and pyroxene (10\%) is consistent with both $B-V$ and $V-R$ colors, and polarization in $R$-band. Such a mixture is also consistent with the observed radial profiles in Fig. \ref{fig:color_radius}. For pyroxene, $B-V$ color increases with size while $V-R$ color remains relatively constant as seen in Fig. \ref{fig:color_model}, indicating that the expected decrease of particle size with the cometocentric distance is consistent with the observed color profiles. It is important to note that, since the refractive indices of organic matter and silicate minerals in comets have large uncertainties, the retrieved composition should only be considered as a model parameter or as a rough estimate of the real composition. In addition, if monomers within aggregates consist of  different components, linear mixing is no longer applicable.

\begin{figure}
    \centering
     \includegraphics[width=0.48\linewidth]{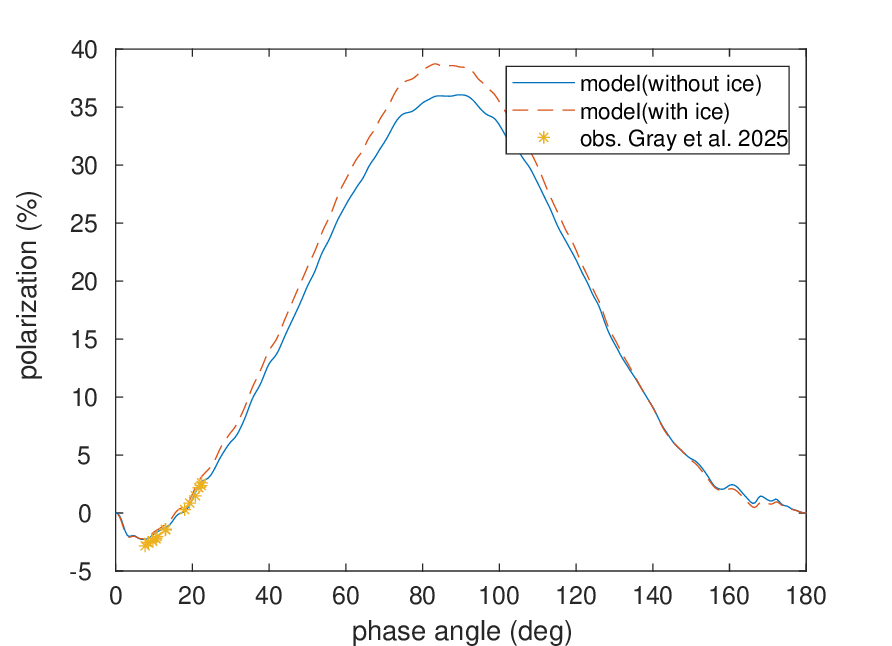}
    \includegraphics[width=0.48\linewidth]{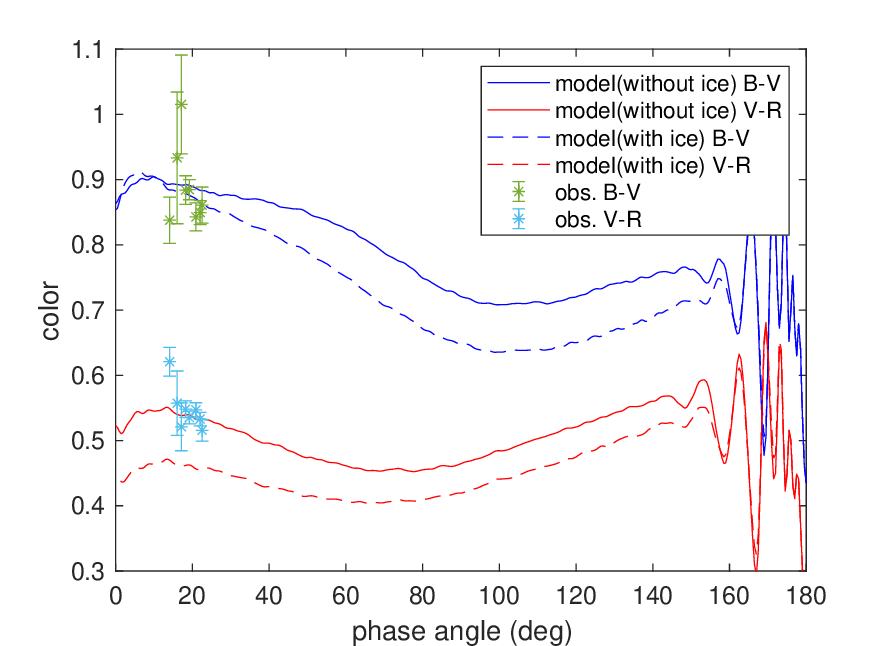}
    \caption{Modeled $R$-band linear polarization and $B-V$ and $V-R$ colors for a mixture of particles with and without 10\% water ice.}
\label{fig:model}
\end{figure}

Since the comet passed through the water ice line, it is interesting to study how sensitive the modeling results are to the ice content. Fig. \ref{fig:model} shows the degree of linear polarization and colors for particles with and without 10\% water ice. It is evident that water ice content has an impact on the colors, which would be detectable if the water ice content changed significantly as the comet approaches the Sun. However, it should be noted that if water ice is mixed within dust aggregates on a micrometer scale, the effect of the ice content on color is negligible.

\section{Conclusions}
\label{sec:conclusions}

Our multi-facility photometric and spectroscopic monitoring of 3I during its pre-perihelion phase reveals a coma significantly dominated by its dust component. The observed red color indices are mutually consistent across our datasets and directly correspond to the slope of the reflected solar continuum, as confirmed by our spectroscopic results. We investigated the temporal evolution of the spectral slope between two observing epochs. Previously published spectra \citep{Santana-Ros2025} exhibited continuous reddening, with slopes increasing from $\sim$15 to $\sim$30 $\%/1000\,\mbox{\AA}$, while the latest data, together with emission appearing over the same wavelength interval, indicate a slight decrease in spectral slope. These spectra are largely characterized by a featureless dust continuum, with CN being the only clearly identified emission feature. Importantly, our analysis shows that the contribution of gaseous emissions to the broadband photometry remains negligible throughout the observed period. The lack of significant temporal variation in these color indices suggests that the dust properties remained relatively constant over the interval, while the detected weak radial color dependence—where the inner coma appears redder than the outer regions—likely reflects a variation in the grain size distribution with cometocentric distance.

These stable dust properties are mirrored in the object's large-scale morphology. By applying digital filtering techniques, including rotational gradient and Laplacian methods, we identified a persistent sunward fan-shaped coma originating from a stable, high-latitude active region on the nucleus. The stability of this structure over several months, coupled with a periodic modulation of the jet position angle, is consistent with the precessional motion of the spin axis as suggested by \citet{2026A&A...705L...3S}. Such a well-defined and persistent activity pattern distinguishes 3I from the inactive 1I/`Oumuamua, while its specific chemical depletion of carbon-chain molecules \citep{Manzano2025} and CO$_2$-dominated volatile profile \citep{Cordiner2025} sets it apart from the more {``primitive"} 2I/Borisov \citep{Bagnulo2021}.

Ultimately, 3I reinforces the emerging picture of vast diversity among the interstellar population. While its dust-dominated coma and taxonomic colors align with many Jupiter-family comets in our own solar system, its unique volatile ratios provide a rare window into the distinct chemical conditions of its parent system. The transition of 3I through perihelion offers a critical baseline for understanding how interstellar objects evolve under solar influence. Continued post-perihelion monitoring will be vital to determine whether the observed chemical depletion and dust characteristics are representative of the bulk nucleus or indicative of surface processing during its journey through the interstellar medium.

\begin{acknowledgments}
T.S-R.\ acknowledges funding from Ministerio de Ciencia e Innovaci{\'o}n (Spanish Government), PGC2021, PID2021-125883NB-C21. This work was (partially) supported by the Spanish MICIN/AEI/10.13039/501100011033 and by ''ERDF A way of making Europe'' by the ''European Union'' through grant PID2021-122842OB-C21, and the Institute of Cosmos Sciences University of Barcelona (ICCUB, Unidad de Excelencia `Mar\'ia de Maeztu’) through grant CEX2019-000918-M. D.O.\ was supported by grant No.\ 2022/45/B/ST9/00267 from the National Science Centre, Poland. 
 Some of the observations reported in this paper were obtained with the Southern African Large Telescope (SALT) under program 2025-1-DDT-004 (PI: Tomasz Kwiatkowski). Polish participation in SALT is funded by grant No.\ MEiN nr 2021/WK/01. Partly based on observations made with the Nordic Optical Telescope. The NOT data were obtained under program ID 71-411 (PI A. Penttil\"a).

This paper is based on observations made with the MuSCAT instruments, developed by the Astrobiology Center (ABC) in Japan, the University of Tokyo, and Las Cumbres Observatory (LCOGT). MuSCAT3 was developed with financial support from JSPS KAKENHI (JP18H05439) and JST PRESTO (JPMJPR1775), and is located at the Faulkes Telescope North on Maui, HI (USA), operated by LCOGT. MuSCAT4 was developed with financial support provided by the Heising-Simons Foundation (grant 2022-3611), JST grant number JPMJCR1761, and the ABC in Japan, and is located at the Faulkes Telescope South at Siding Spring Observatory (Australia), operated by LCOGT.

Some observations were obtained by the Comet Chasers school outreach program (https://www.cometchasers.org/ ), led by Helen Usher, which is funded by the UK Science and Technology Facilities Council (via the DeepSpace2DeepImpact Project), the Open University and Cardiff University. It accesses the LCOGT telescopes through the Schools Observatory/Faulkes Telescope Project (TSO2025A-00 DFET-The Schools’ Observatory), which is partly funded by the Dill Faulkes Educational Trust, and through the LCO Global Sky Partners Program (LCOEPO2023B-013). Observers included E. Maciulis, A. Trawicki, L. Garrett, participants on the British Astronomical Association's Work Experience project 2025, and representatives from the following schools and clubs: Institut d'Alcarràs, Catalonia, Spain; The Coopers' Company \& Coborn School, Upminster, UK; Ysgol Gyfun Gymraeg Bro Edern, Cardiff, UK; St Marys Catholic Primary School, Bridgend, UK;  and Louis Cruls Astronomy Club, Brazil.

The research of OI and AS is supported by the Slovakia-France Grant no. SK-FR-24-0014 and the Project no. 53632UB - PHC STEFANIK 2025. The research of OI is supported by the Slovak Research and Development Agency under Contracts No. APVV-24-0076  and by the Slovak Academy of Sciences (grant Vega No. 2/0067/26). OI, JM are supported by grants for projects under the Mobility Program DAAD-SAS-2024-02 no. 57752921. 
JM\ was supported by the German Research Foundation (DFG) grant no. 517146316. This work used the supercomputer Phoenix and was supported by the Gauß-IT-Zentrum of the University of Braunschweig (GITZ).
YK and AS thank the French PAUSE program for its support of scientists at risk. The authors express their gratitude to
all those people who defend Ukraine and thus made it possible to prepare this article.
\end{acknowledgments}

\begin{contribution}

\textbf{TSR}: Photometric observations with the Faulkes Telescopes and writing the paper. 
\textbf{AS}: Photometric analysis and writing the paper.
\textbf{OI}: Morphology and writing the paper.
\textbf{SM}: Spectral analysis and writing the paper. 
\textbf{JM}: Light-scattering modeling and writing the paper. 
\textbf{IL}: Spectra analysis and writing the paper. 
\textbf{TK}: Campaign coordination and writing the paper. 
\textbf{DO}: NOT spectra observations. 
\textbf{AP}: NOT spectra observations. 
\textbf{NE}: SALT spectra observations. 
\textbf{KE}: Maidanak photometric observations. 
\textbf{YK}: Photometric analysis.
All authors contributed to the discussion of the results and commented on the final manuscript.

\end{contribution}

\facilities{SALT, NOT, LCOGT, Maidanak:1.5m.}

\software{astropy \citep{astropy}, 
          photutils \citep{bradley_2026}, 
          IRAF \citep[see \url{https://iraf-community.github.io/}]{IRAF}, 
          IDL \citep{IDL}, 
          FaSTMM2 \citep{Markkanen2026},
          MuSCAT pipeline \citep[see \url{https://github.com/hpparvi/MuSCAT2_transit_pipeline}]{Narita2020}.
}

\bibliography{sample701}{}

@ARTICLE{Ahearn1995,
       author = {{A'Hearn}, Michael F. and {Millis}, Robert C. and {Schleicher}, David O. and {Osip}, David J. and {Birch}, Peter V.},
        title = "{The Ensemble Properties of Comets: Results from Narrowband Photometry of 85 Comets, 1976-1992}",
      journal = {\icarus},
         year = "1995",
       volume = {118},
        pages = {223-270},
          doi = {10.1006/icar.1995.1190}
}

@ARTICLE{astropy,
       author = {{Astropy Collaboration} and {Price-Whelan}, Adrian M. and {Lim}, Pey Lian and {Earl}, Nicholas and {Starkman}, Nathaniel and {Bradley}, Larry and {Shupe}, David L. and {Patil}, Aarya A. and {Corrales}, Lia and {Brasseur}, C.~E. and {N{\"o}the}, Maximilian and {Donath}, Axel and {Tollerud}, Erik and {Morris}, Brett M. and {Ginsburg}, Adam and {Vaher}, Eero and {Weaver}, Benjamin A. and {Tocknell}, James and {Jamieson}, William and {van Kerkwijk}, Marten H. and {Robitaille}, Thomas P. and {Merry}, Bruce and {Bachetti}, Matteo and {G{\"u}nther}, H. Moritz and {Aldcroft}, Thomas L. and {Alvarado-Montes}, Jaime A. and {Archibald}, Anne M. and {B{\'o}di}, Attila and {Bapat}, Shreyas and {Barentsen}, Geert and {Baz{\'a}n}, Juanjo and {Biswas}, Manish and {Boquien}, M{\'e}d{\'e}ric and {Burke}, D.~J. and {Cara}, Daria and {Cara}, Mihai and {Conroy}, Kyle E. and {Conseil}, Simon and {Craig}, Matthew W. and {Cross}, Robert M. and {Cruz}, Kelle L. and {D'Eugenio}, Francesco and {Dencheva}, Nadia and {Devillepoix}, Hadrien A.~R. and {Dietrich}, J{\"o}rg P. and {Eigenbrot}, Arthur Davis and {Erben}, Thomas and {Ferreira}, Leonardo and {Foreman-Mackey}, Daniel and {Fox}, Ryan and {Freij}, Nabil and {Garg}, Suyog and {Geda}, Robel and {Glattly}, Lauren and {Gondhalekar}, Yash and {Gordon}, Karl D. and {Grant}, David and {Greenfield}, Perry and {Groener}, Austen M. and {Guest}, Steve and {Gurovich}, Sebastian and {Handberg}, Rasmus and {Hart}, Akeem and {Hatfield-Dodds}, Zac and {Homeier}, Derek and {Hosseinzadeh}, Griffin and {Jenness}, Tim and {Jones}, Craig K. and {Joseph}, Prajwel and {Kalmbach}, J. Bryce and {Karamehmetoglu}, Emir and {Ka{\l}uszy{\'n}ski}, Miko{\l}aj and {Kelley}, Michael S.~P. and {Kern}, Nicholas and {Kerzendorf}, Wolfgang E. and {Koch}, Eric W. and {Kulumani}, Shankar and {Lee}, Antony and {Ly}, Chun and {Ma}, Zhiyuan and {MacBride}, Conor and {Maljaars}, Jakob M. and {Muna}, Demitri and {Murphy}, N.~A. and {Norman}, Henrik and {O'Steen}, Richard and {Oman}, Kyle A. and {Pacifici}, Camilla and {Pascual}, Sergio and {Pascual-Granado}, J. and {Patil}, Rohit R. and {Perren}, Gabriel I. and {Pickering}, Timothy E. and {Rastogi}, Tanuj and {Roulston}, Benjamin R. and {Ryan}, Daniel F. and {Rykoff}, Eli S. and {Sabater}, Jose and {Sakurikar}, Parikshit and {Salgado}, Jes{\'u}s and {Sanghi}, Aniket and {Saunders}, Nicholas and {Savchenko}, Volodymyr and {Schwardt}, Ludwig and {Seifert-Eckert}, Michael and {Shih}, Albert Y. and {Jain}, Anany Shrey and {Shukla}, Gyanendra and {Sick}, Jonathan and {Simpson}, Chris and {Singanamalla}, Sudheesh and {Singer}, Leo P. and {Singhal}, Jaladh and {Sinha}, Manodeep and {Sip{\H{o}}cz}, Brigitta M. and {Spitler}, Lee R. and {Stansby}, David and {Streicher}, Ole and {{\v{S}}umak}, Jani and {Swinbank}, John D. and {Taranu}, Dan S. and {Tewary}, Nikita and {Tremblay}, Grant R. and {Val-Borro}, Miguel de and {Van Kooten}, Samuel J. and {Vasovi{\'c}}, Zlatan and {Verma}, Shresth and {de Miranda Cardoso}, Jos{\'e} Vin{\'i}cius and {Williams}, Peter K.~G. and {Wilson}, Tom J. and {Winkel}, Benjamin and {Wood-Vasey}, W.~M. and {Xue}, Rui and {Yoachim}, Peter and {Zhang}, Chen and {Zonca}, Andrea and {Astropy Project Contributors}},
       title = "{The Astropy Project: Sustaining and Growing a Community-oriented Open-source Project and the Latest Major Release (v5.0) of the Core Package}",
     journal = {\apj},
        year = 2022,
       volume = {935},
         eid = {167},
         doi = {10.3847/1538-4357/ac7c74},
      adsurl = {https://ui.adsabs.harvard.edu/abs/2022ApJ...935..167A},
}

@ARTICLE{Bagnulo2021,
       author = {{Bagnulo}, S. and {Cellino}, A. and {Kolokolova}, L. and {Ne{\v{z}}i{\v{c}}}, R. and {Santana-Ros}, T. and {Borisov}, G. and {Christou}, A.~A. and {Bendjoya}, Ph. and {Devog{\`e}le}, M.},   
        title = "{Unusual polarimetric properties of the first interstellar comet 2I/Borisov}",
      journal = {Nature Communications},
         year = "2021",
       volume = {12},
        pages = {1797},
          doi = {10.1038/s41467-021-22000-x}
}

@ARTICLE{Beniyama2025,
       author = {{Beniyama}, Jin},
        title = "{Simultaneous visible spectrophotometry of interstellar object 3I/ATLAS with Seimei/TriCCS}",
      journal = {\pasj},
         year = 2025,
        month = oct,
       volume = {77},
       number = {5},
        pages = {L71-L76},
          doi = {10.1093/pasj/psaf097},
archivePrefix = {arXiv},
       eprint = {2508.08829},
 primaryClass = {astro-ph.EP},
       adsurl = {https://ui.adsabs.harvard.edu/abs/2025PASJ...77L..71B}
}

@ARTICLE{Bolin25,
       author = {{Bolin}, Bryce T. and {Belyakov}, Matthew and {Fremling}, Christoffer and {Graham}, Matthew J. and {Abdelaziz}, Ahmed M. and {Elhosseiny}, Eslam and {Gray}, Candace L. and {Ingebretsen}, Carl and {Jewett}, Gracyn and {Lisse}, Carey M. and {Karpov}, Sergey and {Kilic}, Mukremin and {Ma{\v{s}}ek}, Martin and {Molham}, Mona and {Roderick}, Diana and {Takey}, Ali and {Abron}, Laura-May and {Coughlin}, Michael W. and {Hsieh}, Cheng-Han and {Noll}, Keith S. and {Wong}, Ian},
        title = "{Interstellar comet 3I/ATLAS: discovery and physical description}",
      journal = {\mnras},
         year = 2025,
        month = sep,
       volume = {542},
       number = {1},
        pages = {L139-L143},
          doi = {10.1093/mnrasl/slaf078},
archivePrefix = {arXiv},
       eprint = {2507.05252},
 primaryClass = {astro-ph.EP},
       adsurl = {https://ui.adsabs.harvard.edu/abs/2025MNRAS.542L.139B}
}

@ARTICLE{2025ATel17363....1B,
       author = {{Bolin}, Bryce and {Abron}, Laura-May and {Belyakov}, Matthew and {Carvajal}, Juan Pablo and {Fremling}, Christoffer and {Hanus}, Josef and {Ingebretsen}, Carl and {King}, Piera and {Luco}, Baltasar and {Noll}, Keith and {Puzia}, Thomas H. and {Rahatgaonkar}, Rohan and {Silva}, Karleyne and {Wong}, Ian},
       title = "{The Anti-Solar Tail of Comet 3I/ATLAS Seen in Deep Multi-band Imaging Taken at Gemini South}",
      journal = {The Astronomer's Telegram},
         year = "2025",
       volume = {17363},
        pages = {1}
}

@MISC{bradley_2026,
author = {{Bradley}, Larry and {Sip{\H{o}}cz}, Brigitta M. and {Robitaille}, T. P. and {Tollerud}, E. J. and {Vin{\'\i}cius}, Z{\'e} and {Deil}, Christoph and {Barbary}, Kyle and {Wilson}, Tom J. and {Busko}, Ivo and {Donath}, Axel and {G{\"u}nther}, Hans Moritz and {Cara}, Mihai and {Lim}, P. L. and {Me{\ss}linger}, Sebastian and {Conseil}, Simon and {Droettboom}, Michael and {Bostroem}, K. Azalee and {Bray}, E. M. and {Bratholm}, Lars Andersen and {Burnett}, Zach and {Jamieson}, William and {Ginsburg}, Adam and {Taranu}, Dan and {Barentsen}, Geert and {Craig}, Matthew W. and {Morris}, Brett M. and {Perrin}, Marshall and {Rathi}, Shivangee}, 
title = "{Photutils}", 
year = {2026}, 
month = apr, 
publisher = {Zenodo}, 
version = {3.0.0}, 
doi = {10.5281/zenodo.19636730}, 
url = {https://doi.org/10.5281/zenodo.19636730} 
}

@ARTICLE{Cordiner2025,
       author = {{Cordiner}, Martin A. and {Roth}, Nathan X. and {Kelley}, Michael S.~P. and {Bodewits}, Dennis and {Charnley}, Steven B. and {Drozdovskaya}, Maria N. and {Farnocchia}, Davide and {Micheli}, Marco and {Milam}, Stefanie N. and {Opitom}, Cyrielle and {Schwamb}, Megan E. and {Thomas}, Cristina A. and {Bagnulo}, Stefano},
        title = "{JWST Detection of a Carbon-dioxide-dominated Gas Coma Surrounding Interstellar Object 3I/ATLAS}",
      journal = {\apjl},
         year = 2025,
        month = oct,
       volume = {991},
       number = {2},
          eid = {L43},
        pages = {L43},
          doi = {10.3847/2041-8213/ae0647},
archivePrefix = {arXiv},
       eprint = {2508.18209},
 primaryClass = {astro-ph.EP},
       adsurl = {https://ui.adsabs.harvard.edu/abs/2025ApJ...991L..43C}
}

@ARTICLE{deLaFuenteMarcos2025,
       author = {{de la Fuente Marcos}, R. and {Alarcon}, M.~R. and {Licandro}, J. and {Serra-Ricart}, M. and {de Le{\'o}n}, J. and {de la Fuente Marcos}, C. and {Lombardi}, G. and {Tejero}, A. and {Cabrera-Lavers}, A. and {Guerra Arencibia}, S. and {Ruiz Cejudo}, I.},  
      journal = {\aap},
         year = "2025",
       volume = {700},
        pages = {L9},
          doi = {10.1051/0004-6361/202556439}
}

@ARTICLE{Dorschner1995,
       author = {{Dorschner}, J. and {Begemann}, B. and {Henning}, T. and {Jaeger}, C. and {Mutschke}, H.},
      journal = {\aap},
         year = "1995",
       volume = {300},
        pages = {503}
}

@ARTICLE{Farnham2009,
       author = {{Farnham}, T. L.},
      journal = {\planss},
         year = "2009",
       volume = {57},
        pages = {1192-1217}
}

@ARTICLE{Fukui2022,
       author = {{Fukui}, Akihiko and {Kimura}, Tadahiro and {Hirano}, Teruyuki and {Narita}, Norio and {Kodama}, Takanori and {Hori}, Yasunori and {Ikoma}, Masahiro and {Pall{\'e}}, Enric and {Murgas}, Felipe and {Parviainen}, Hannu and {Kawauchi}, Kiyoe and {Mori}, Mayuko and {Esparza-Borges}, Emma and {Bieryla}, Allyson and {Irwin}, Jonathan and {Safonov}, Boris S. and {Stassun}, Keivan G. and {Alvarez-Hernandez}, Leticia and {B{\'e}jar}, V{\'\i}ctor J.~S. and {Casasayas-Barris}, N{\'u}ria and {Chen}, Guo and {Crouzet}, Nicolas and {de Leon}, Jerome P. and {Isogai}, Keisuke and {Kagetani}, Taiki and {Klagyivik}, Peter and {Korth}, Judith and {Kurita}, Seiya and {Kusakabe}, Nobuhiko and {Livingston}, John and {Luque}, Rafael and {Madrigal-Aguado}, Alberto and {Morello}, Giuseppe and {Nishiumi}, Taku and {Orell-Miquel}, Jaume and {Oshagh}, Mahmoudreza and {S{\'a}nchez-Benavente}, Manuel and {Stangret}, Monika and {Terada}, Yuka and {Watanabe}, Noriharu and {Zou}, Yujie and {Tamura}, Motohide and {Kurokawa}, Takashi and {Kuzuhara}, Masayuki and {Nishikawa}, Jun and {Omiya}, Masashi and {Vievard}, S{\'e}bastien and {Ueda}, Akitoshi and {Latham}, David W. and {Quinn}, Samuel N. and {Strakhov}, Ivan S. and {Belinski}, Alexandr A. and {Jenkins}, Jon M. and {Ricker}, George R. and {Seager}, Sara and {Vanderspek}, Roland and {Winn}, Joshua N. and {Charbonneau}, David and {Ciardi}, David R. and {Collins}, Karen A. and {Doty}, John P. and {Bachelet}, Etienne and {Harbeck}, Daniel},
        title = "{TOI-2285b: A 1.7 Earth-radius planet near the habitable zone around a nearby M dwarf}",
      journal = {\pasj},
         year = 2022,
        month = feb,
       volume = {74},
       number = {1},
        pages = {L1-L8},
          doi = {10.1093/pasj/psab106},
archivePrefix = {arXiv},
       eprint = {2110.10215},
 primaryClass = {astro-ph.EP},
       adsurl = {https://ui.adsabs.harvard.edu/abs/2022PASJ...74L...1F}
}

@INPROCEEDINGS{Giorgini1996,
       author = {{Giorgini}, J.~D. and {Yeomans}, D.~K. and {Chamberlin}, A.~B. and {Chodas}, P.~W. and {Jacobson}, R.~A. and {Keesey}, M.~S. and {Lieske}, J.~H. and {Ostro}, S.~J. and {Standish}, E.~M. and {Wimberly}, R.~N.},
        title = "{JPL's On-Line Solar System Data Service}",
    booktitle = {AAS/Division for Planetary Sciences Meeting Abstracts \#28},
         year = 1996,
       series = {AAS/Division for Planetary Sciences Meeting Abstracts},
       volume = {28},
        month = sep,
          eid = {25.04},
        pages = {25.04},
       adsurl = {https://ui.adsabs.harvard.edu/abs/1996DPS....28.2504G}
}

@ARTICLE{Ginsburg2019,
       author = {{Ginsburg}, Adam and {Sip{\H{o}}cz}, Brigitta M. and {Brasseur}, C. E. and {Cowperthwaite}, Philip S. and {Craig}, Matthew W. and {Deil}, Christoph and {Guillochon}, James and {Guzman}, Giannina and {Liedtke}, Simon and {Lim}, Pey Lian and {Shupe}, David L. and {Bostroem}, K. Azalee and {Stansby}, David and {Droettboom}, Michael and {Tan}, Tan S. and {Lian-Yew}, Chris and {Lu}, Jialin and {Kerzendorf}, Wolfgang and {Astroquery Contributors}},
        title = "{Astroquery: An Astronomical Web-querying Package in Python}",
      journal = {\aj},
         year = "2019",
       volume = {157},
        pages = {98},
          doi = {10.3847/1538-3881/aafc33}
}

@ARTICLE{Gray2025,
       author = {{Gray}, Zuri and {Bagnulo}, Stefano and {Borisov}, Galin and {Kwon}, Yuna G. and {Cellino}, Alberto and {Kolokolova}, Ludmilla and {Dorsey}, Rosemary C. and {Fedorets}, Grigori and {Granvik}, Mikael and {MacLennan}, Eric and {Mu{\~n}oz}, Olga and {Bendjoya}, Philippe and {Devog{\`e}le}, Maxime and {Ieva}, Simone and {Penttil{\"a}}, Antti and {Muinonen}, Karri},
        title = "{Extreme Negative Polarization of New Interstellar Comet 3I/ATLAS}",
      journal = {\apjl},
         year = 2025,
        month = oct,
       volume = {992},
       number = {2},
          eid = {L29},
        pages = {L29},
          doi = {10.3847/2041-8213/ae0c08},
archivePrefix = {arXiv},
       eprint = {2509.05181},
 primaryClass = {astro-ph.EP},
       adsurl = {https://ui.adsabs.harvard.edu/abs/2025ApJ...992L..29G}
}

@ARTICLE{haser1957,
       author = {{Haser}, L.},
        title = "{Distribution d'intensit{\'e} dans la t{\^e}te d'une com{\`e}te}",
      journal = {Bulletin de la Societe Royale des Sciences de Liege},
         year = 1957,
        month = jan,
       volume = {43},
        pages = {740-750},
       adsurl = {https://ui.adsabs.harvard.edu/abs/1957BSRSL..43..740H}
}

@ARTICLE{2025ATel17372....1I,
       author = {{Ivanova}, Oleksandra and {Kwiatkowski}, Tomasz and {Erasmus}, Nicolas and {Mykhailova}, Sofiia and {Oszkiewicz}, Dagmara and {Santana-Ros}, Tony and {Ngwane}, Thobekile and {Penttila}, Antti and {Husaric}, Marek and {Kaminski}, Krzysztof},
        title = "{Follow-up observations of the dust fan ejected from interstellar comet 3I/ATLAS}",
      journal = {The Astronomer's Telegram},
         year = 2025,
        month = sep,
       volume = {17372},
        pages = {1},
       adsurl = {https://ui.adsabs.harvard.edu/abs/2025ATel17372....1I}
}

@ARTICLE{jenniskens1993,
       author = {{Jenniskens}, P.},
      journal = {\aap},
         year = "1993",
       volume = {274},
        pages = {653}
}

@ARTICLE{2025ApJ...994L...3J,
       author = {{Jewitt}, David and {Luu}, Jane},
        title = "{Preperihelion Development of Interstellar Comet 3I/ATLAS}",
      journal = {\apjl},
         year = 2025,
        month = nov,
       volume = {994},
       number = {1},
          eid = {L3},
        pages = {L3},
          doi = {10.3847/2041-8213/ae1832},
archivePrefix = {arXiv},
       eprint = {2510.18769},
 primaryClass = {astro-ph.EP},
       adsurl = {https://ui.adsabs.harvard.edu/abs/2025ApJ...994L...3J}
}

@ARTICLE{Kareta2025,
       author = {{Kareta}, Theodore and {Champagne}, Chansey and {McClure}, Lucas and {Emery}, Joshua and {Sharkey}, Benjamin N.~L. and {Bauer}, James and {Connelley}, Michael S. and {Rayner}, John and {Thomas}, Cristina A. and {Reddy}, Vishnu and {Firgard}, Megan},
        title = "{Near-discovery Observations of Interstellar Comet 3I/ATLAS with the NASA Infrared Telescope Facility}",
      journal = {\apjl},
         year = 2025,
        month = sep,
       volume = {990},
       number = {2},
          eid = {L65},
        pages = {L65},
          doi = {10.3847/2041-8213/adfbdf},
archivePrefix = {arXiv},
       eprint = {2507.12234},
 primaryClass = {astro-ph.EP},
       adsurl = {https://ui.adsabs.harvard.edu/abs/2025ApJ...990L..65K}
}

@ARTICLE{Kim1994,
       author = {{Kim}, Sang J.},
        title = "{Ultraviolet and Visible Spectroscopic Database for Atoms and Molecules in Celestial Objects}",
      journal = {Publication of Korean Astronomical Society},
         year = 1994,
        month = dec,
       volume = {9},
        pages = {111-166},
       adsurl = {https://ui.adsabs.harvard.edu/abs/1994PKAS....9..111K}
}

@INPROCEEDINGS{IDL,
       author = {{Landsman}, W.~B.},
        title = "{The IDL Astronomy User's Library}",
    booktitle = {Astronomical Data Analysis Software and Systems IV},
         year = 1995,
       editor = {{Shaw}, R.~A. and {Payne}, H.~E. and {Hayes}, J.~J.~E.},
       series = {Astronomical Society of the Pacific Conference Series},
       volume = {77},
        month = jan,
        pages = {437},
       adsurl = {https://ui.adsabs.harvard.edu/abs/1995ASPC...77..437L}
}

@ARTICLE{2011Icar..213..280L,
       author = {{Langland-Shula}, L. E. and {Smith}, G. H.},
      journal = {\icarus},
         year = "2011",
       volume = {213},
        pages = {280}
}

@ARTICLE{larson1984coma,
       author = {{Larson}, S. and {Sekanina}, Z.},
      journal = {\aj},
         year = "1984",
       volume = {89},
        pages = {571}
}

@ARTICLE{2020ApJS..251....6M,
       author = {{Magnier}, Eugene. A. and {Schlafly}, Edward. F. and {Finkbeiner}, Douglas P. and {Tonry}, J.~L. and {Goldman}, B. and {R{\"o}ser}, S. and {Schilbach}, E. and {Casertano}, S. and {Chambers}, K.~C. and {Flewelling}, H.~A. and {Huber}, M.~E. and {Price}, P.~A. and {Sweeney}, W.~E. and {Waters}, C.~Z. and {Denneau}, L. and {Draper}, P.~W. and {Hodapp}, K.~W. and {Jedicke}, R. and {Kaiser}, N. and {Kudritzki}, R.-P. and {Metcalfe}, N. and {Stubbs}, C.~W. and {Wainscoat}, R.~J.},
        title = "{Pan-STARRS Photometric and Astrometric Calibration}",
      journal = {\apjs},
         year = 2020,
        month = nov,
       volume = {251},
       number = {1},
          eid = {6},
        pages = {6},
          doi = {10.3847/1538-4365/abb82a},
archivePrefix = {arXiv},
       eprint = {1612.05242},
 primaryClass = {astro-ph.IM},
       adsurl = {https://ui.adsabs.harvard.edu/abs/2020ApJS..251....6M}
}

@ARTICLE{Markkanen2017,
       author = {{Markkanen}, J. and {Yuffa}, A.},
      journal = {JQSRT},
         year = "2017",
       volume = {189},
        pages = {181},
        doi = {10.1016/j.jqsrt.2016.11.004}
}

@MISC{Markkanen2026,
       author = {{Markkanen}, J.},
        title = "{Fast superposition T-matrix software FaSTMM2}",
         year = "2026",
howpublished = {Zenodo},
          doi = {10.5281/zenodo.18412809}
}

@INPROCEEDINGS{Narita2020,
       author = {{Narita}, Norio and {Fukui}, Akihiko and {Yamamuro}, Tomoyasu and
                 {Harbeck}, Daniel and {Bowman}, Mark and {Elphick}, Mark and
                 {Nation}, Jon and {Armstrong}, J.~D. and {Han}, Jacqueline and
                 {Abe}, Shunichi and {Ikoma}, Masahiro and {Isogai}, Keisuke and
                 {Kawauchi}, Kiyoe and {Kurita}, Seiya and {Kusakabe}, Nobuhiko and
                 {de Leon}, Jerome and {Livingston}, John and {Mori}, Mayuko and
                 {Nishiumi}, Taku and {Tamura}, Motohide and {Watanabe}, Noriharu and
                 {Volgenau}, Nikolaus and {Heinrich-Josties}, Elisabeth and
                 {Foale}, Steve and {Daily}, Matt and {McCully}, Curtis and
                 {Kirby}, Annie and {Smith}, Cary and {Haworth}, Brian and
                 {Conway}, Patrick and {Storrie-Lombardi}, Lisa and {Rosing}, Wayne and
                 {Chatelain}, Joey and {Bachelet}, Etienne and {Johnson}, Marshall and
                 {Rabus}, Markus},
        title = "{MuSCAT3: a 4-color simultaneous camera for the 2m Faulkes Telescope North}",
    booktitle = {Ground-based and Airborne Instrumentation for Astronomy VIII},
       series = {Society of Photo-Optical Instrumentation Engineers (SPIE) Conference Series},
         year = {2020},
       volume = {11447},
        pages = {114475K},
          eid = {114475K},
          doi = {10.1117/12.2559947}
}

@ARTICLE{2025ATel17350....1O,
       author = {{Oldani}, Virginio and {Manzini}, Federico and {Reguitti}, Andrea and
                 {Mura}, Alessandra and {Ochner}, Paolo and {Bedin}, Luigi R. and
                 {Farina}, Andrea},
        title = "{Fan-shaped dust emission from the nucleus of comet 3I}",
      journal = {The Astronomer's Telegram},
         year = {2025},
        month = aug,
       volume = {17350},
        pages = {1},
       adsurl = {https://ui.adsabs.harvard.edu/abs/2025ATel17350....1O}
}

@ARTICLE{Manzano2025,
       author = {{Salazar Manzano}, Luis E. and {Lin}, Hsing Wen and
                 {Taylor}, Aster G. and {Seligman}, Darryl Z. and
                 {Adams}, Fred C. and {Gerdes}, David W. and
                 {Ruch}, Thomas and {Frincke}, Tessa T. and
                 {Napier}, Kevin J.},
        title = "{Onset of CN Emission in 3I/ATLAS: Evidence for Strong Carbon-chain Depletion}",
      journal = {\apjl},
         year = {2025},
       volume = {993},
       number = {1},
        pages = {L23},
          eid = {L23},
          doi = {10.3847/2041-8213/ae1232},
archivePrefix = {arXiv},
       eprint = {2509.01647},
 primaryClass = {astro-ph.EP},
       adsurl = {https://ui.adsabs.harvard.edu/abs/2025ApJ...993L..23S}
}

@ARTICLE{samarasinha2014image,
       author = {{Samarasinha}, Nalin H. and {Larson}, Stephen M.},
        title = "{Image enhancement techniques for quantitative investigations of morphological features in cometary comae: A comparative study}",
      journal = {\icarus},
         year = 2014,
        month = sep,
       volume = {239},
        pages = {168-185},
          doi = {10.1016/j.icarus.2014.05.028},
archivePrefix = {arXiv},
       eprint = {1406.0033},
 primaryClass = {astro-ph.EP},
       adsurl = {https://ui.adsabs.harvard.edu/abs/2014Icar..239..168S}
}

@ARTICLE{Santana-Ros2025,
       author = {{Santana-Ros}, T. and {Ivanova}, O. and {Mykhailova}, S. and {Erasmus}, N. and {Kami{\'n}ski}, K. and {Oszkiewicz}, D. and {Kwiatkowski}, T. and {Hus{\'a}rik}, M. and {Ngwane}, T.~S. and {Penttil{\"a}}, A.},
        title = "{Temporal evolution of the third interstellar comet 3I/ATLAS: Spin, color, spectra, and dust activity}",
      journal = {\aap},
         year = 2025,
        month = oct,
       volume = {702},
          eid = {L3},
        pages = {L3},
          doi = {10.1051/0004-6361/202556717},
archivePrefix = {arXiv},
       eprint = {2508.00808},
 primaryClass = {astro-ph.EP},
       adsurl = {https://ui.adsabs.harvard.edu/abs/2025A&A...702L...3S}
}

@ARTICLE{Schleicher1993,
       author = {{Schleicher}, David G. and {Bus}, Schelte J. and {Osip}, David J.},
        title = "{The Anomalous Molecular Abundances of Comet P/Wolf-Harrington}",
      journal = {\icarus},
         year = 1993,
        month = aug,
       volume = {104},
       number = {2},
        pages = {157-166},
          doi = {10.1006/icar.1993.1092},
       adsurl = {https://ui.adsabs.harvard.edu/abs/1993Icar..104..157S}
}

@ARTICLE{2010AJ....140..973S,
       author = {{Schleicher}, David G.},
        title = "{The Fluorescence Efficiencies of the CN Violet Bands in Comets}",
      journal = {\aj},
         year = 2010,
        month = oct,
       volume = {140},
       number = {4},
        pages = {973-984},
          doi = {10.1088/0004-6256/140/4/973},
       adsurl = {https://ui.adsabs.harvard.edu/abs/2010AJ....140..973S}
}

@ARTICLE{Schleicher2025,
       author = {{Schleicher}, D.},
        title = "{The Detection of CN in Interstellar Comet 3I/ATLAS}",
      journal = {The Astronomer's Telegram},
         year = "2025",
       volume = {17352},
        pages = {1}
}

@ARTICLE{Seligman25,
       author = {{Seligman}, Darryl Z. and {Micheli}, Marco and {Farnocchia}, Davide and {Denneau}, Larry and {Noonan}, John W. and {Hsieh}, Henry H. and {Santana-Ros}, Toni and {Tonry}, John and {Auchettl}, Katie and {Conversi}, Luca and {Devog{\`e}le}, Maxime and {Faggioli}, Laura and {Feinstein}, Adina D. and {Fenucci}, Marco and {Ferrais}, Marin and {Frincke}, Tessa and {Gillon}, Michael and {Hainaut}, Olivier R. and {Hart}, Kyle and {Hoffman}, Andrew and {Holt}, Carrie E. and {Hoogendam}, Willem B. and {Huber}, Mark E. and {Jehin}, Emmanuel and {Kareta}, Theodore and {Keane}, Jacqueline V. and {Kelley}, Michael S.~P. and {Lister}, Tim and {Mandt}, Kathleen and {Manfroid}, Jean and {Mar{\v{c}}eta}, Du{\v{s}}an and {Meech}, Karen J. and {Amine Miftah}, Mohamed and {Morgan}, Marvin and {Oca{\~n}a}, Francisco and {Pe{\~n}a-Asensio}, Eloy and {Shappee}, Benjamin J. and {Siverd}, Robert J. and {Taylor}, Aster G. and {Tucker}, Michael A. and {Wainscoat}, Richard and {Weryk}, Robert and {Wray}, James J. and {Yaginuma}, Atsuhiro and {Yang}, Bin and {Ye}, Quanzhi and {Zhang}, Qicheng},
        title = "{Discovery and Preliminary Characterization of a Third Interstellar Object: 3I/ATLAS}",
      journal = {\apjl},
         year = 2025,
        month = aug,
       volume = {989},
       number = {2},
          eid = {L36},
        pages = {L36},
          doi = {10.3847/2041-8213/adf49a},
archivePrefix = {arXiv},
       eprint = {2507.02757},
 primaryClass = {astro-ph.EP},
       adsurl = {https://ui.adsabs.harvard.edu/abs/2025ApJ...989L..36S}
}

@ARTICLE{2026A&A...705L...3S,
       author = {{Serra-Ricart}, M. and {Licandro}, J. and {Alarcon}, M.~R.},
        title = "{Pre-perihelion detection of a wobbling high-latitude jet in the interstellar comet 3I/ATLAS}",
      journal = {\aap},
         year = 2026,
        month = jan,
       volume = {705},
          eid = {L3},
        pages = {L3},
          doi = {10.1051/0004-6361/202558072},
archivePrefix = {arXiv},
       eprint = {2512.12819},
 primaryClass = {astro-ph.EP},
       adsurl = {https://ui.adsabs.harvard.edu/abs/2026A&A...705L...3S}
}

@ARTICLE{Shen2008,
       author = {{Shen}, Yue and {Draine}, B.~T. and {Johnson}, Eric T.},
        title = "{Modeling Porous Dust Grains with Ballistic Aggregates. I. Geometry and Optical Properties}",
      journal = {\apj},
         year = 2008,
        month = dec,
       volume = {689},
       number = {1},
        pages = {260-275},
          doi = {10.1086/592765},
archivePrefix = {arXiv},
       eprint = {0801.1996},
 primaryClass = {astro-ph},
       adsurl = {https://ui.adsabs.harvard.edu/abs/2008ApJ...689..260S}
}

@INPROCEEDINGS{IRAF,
       author = {{Tody}, Doug},
        title = "{The IRAF Data Reduction and Analysis System}",
    booktitle = {Instrumentation in astronomy VI},
         year = 1986,
       editor = {{Crawford}, David L.},
       series = {Society of Photo-Optical Instrumentation Engineers (SPIE) Conference Series},
       volume = {627},
        month = jan,
        pages = {733},
          doi = {10.1117/12.968154},
       adsurl = {https://ui.adsabs.harvard.edu/abs/1986SPIE..627..733T}
}

@ARTICLE{warren1984,
       author = {{Warren}, Stephen G.},
        title = "{Optical constants of ice from the ultraviolet to the microwave}",
      journal = {Applied Optics},
         year = {1984},
        month = apr,
       volume = {23},
       number = {8},
        pages = {1206--1225},
          doi = {10.1364/AO.23.001206},
       adsurl = {https://ui.adsabs.harvard.edu/abs/1984ApOpt..23.1206W}
}

@ARTICLE{Strom2020,
       author = {{Str{\o}m}, Paul A. and {Bodewits}, Dennis and {Knight}, Matthew M. and {Kiefer}, Flavien and {Jones}, Geraint H. and {Kral}, Quentin and {Matr{\`a}}, Luca and {Bodman}, Eva and {Capria}, Maria Teresa and {Cleeves}, Ilsedore and {Fitzsimmons}, Alan and {Haghighipour}, Nader and {Harrison}, John H.~D. and {Iglesias}, Daniela and {Kama}, Mihkel and {Linnartz}, Harold and {Majumdar}, Liton and {de Mooij}, Ernst J.~W. and {Milam}, Stefanie N. and {Opitom}, Cyrielle and {Rebollido}, Isabel and {Rogers}, Laura K. and {Snodgrass}, Colin and {Sousa-Silva}, Clara and {Xu}, Siyi and {Lin}, Zhong-Yi and {Zieba}, Sebastian},
        title = "{Exocomets from a Solar System Perspective}",
      journal = {\pasp},
         year = 2020,
        month = oct,
       volume = {132},
       number = {1016},
          eid = {101001},
        pages = {101001},
          doi = {10.1088/1538-3873/aba6a0},
archivePrefix = {arXiv},
       eprint = {2007.09155},
 primaryClass = {astro-ph.EP},
       adsurl = {https://ui.adsabs.harvard.edu/abs/2020PASP..132j1001S}
}

@ARTICLE{Iglesias2025,
       author = {{Iglesias}, Daniela and {Rebollido}, Isabel and {Norazman}, Azib and {Snodgrass}, Colin and {Seligman}, Darryl Z. and {Xu}, Siyi and {Hoeijmakers}, H. Jens and {Kenworthy}, Matthew and {Lecavelier des Etangs}, Alain and {Bannister}, Michele and {Yang}, Bin},
        title = "{An Overview of Exocomets}",
      journal = {\ssr},
         year = 2025,
        month = dec,
       volume = {221},
       number = {8},
          eid = {122},
        pages = {122},
          doi = {10.1007/s11214-025-01247-6},
archivePrefix = {arXiv},
       eprint = {2511.08270},
 primaryClass = {astro-ph.EP},
       adsurl = {https://ui.adsabs.harvard.edu/abs/2025SSRv..221..122I}
}

@ARTICLE{Kiefer2014,
       author = {{Kiefer}, F. and {Lecavelier des Etangs}, A. and {Boissier}, J. and {Vidal-Madjar}, A. and {Beust}, H. and {Lagrange}, A.-M. and {H{\'e}brard}, G. and {Ferlet}, R.},
        title = "{Two families of exocomets in the {\ensuremath{\beta}} Pictoris system}",
      journal = {\nat},
         year = 2014,
        month = oct,
       volume = {514},
       number = {7523},
        pages = {462-464},
          doi = {10.1038/nature13849},
       adsurl = {https://ui.adsabs.harvard.edu/abs/2014Natur.514..462K}
}

@ARTICLE{Rappaport2018,
       author = {{Rappaport}, S. and {Vanderburg}, A. and {Jacobs}, T. and {LaCourse}, D. and {Jenkins}, J. and {Kraus}, A. and {Rizzuto}, A. and {Latham}, D.~W. and {Bieryla}, A. and {Lazarevic}, M. and {Schmitt}, A.},
        title = "{Likely transiting exocomets detected by Kepler}",
      journal = {\mnras},
         year = 2018,
        month = feb,
       volume = {474},
       number = {2},
        pages = {1453-1468},
          doi = {10.1093/mnras/stx2735},
archivePrefix = {arXiv},
       eprint = {1708.06069},
 primaryClass = {astro-ph.EP},
       adsurl = {https://ui.adsabs.harvard.edu/abs/2018MNRAS.474.1453R}
}

@ARTICLE{Zieba2019,
       author = {{Zieba}, S. and {Zwintz}, K. and {Kenworthy}, M. A. and {Kennedy}, G. M.},
        title = "{Transiting exocomets detected in broadband light by TESS in the {$\beta$} Pictoris system}",
      journal = {\aap},
         year = "2019",
       volume = {625},
        pages = {L13},
          doi = {10.1051/0004-6361/201935552}
}

@ARTICLE{Matra2017betaPic,
       author = {{Matr{\`a}}, L. and {Dent}, W.~R.~F. and {Wyatt}, M.~C. and {Kral}, Q. and {Wilner}, D.~J. and {Pani{\'c}}, O. and {Hughes}, A.~M. and {de Gregorio-Monsalvo}, I. and {Hales}, A. and {Augereau}, J.-C. and {Greaves}, J. and {Roberge}, A.},
        title = "{Exocometary gas structure, origin and physical properties around {\ensuremath{\beta}} Pictoris through ALMA CO multitransition observations}",
      journal = {\mnras},
         year = 2017,
        month = jan,
       volume = {464},
       number = {2},
        pages = {1415-1433},
          doi = {10.1093/mnras/stw2415},
archivePrefix = {arXiv},
       eprint = {1609.06718},
 primaryClass = {astro-ph.EP},
       adsurl = {https://ui.adsabs.harvard.edu/abs/2017MNRAS.464.1415M}
}

@ARTICLE{Matra2017Fomalhaut,
       author = {{Matr{\`a}}, L. and {MacGregor}, M.~A. and {Kalas}, P. and {Wyatt}, M.~C. and {Kennedy}, G.~M. and {Wilner}, D.~J. and {Duchene}, G. and {Hughes}, A.~M. and {Pan}, M. and {Shannon}, A. and {Clampin}, M. and {Fitzgerald}, M.~P. and {Graham}, J.~R. and {Holland}, W.~S. and {Pani{\'c}}, O. and {Su}, K.~Y.~L.},
        title = "{Detection of Exocometary CO within the 440 Myr Old Fomalhaut Belt: A Similar CO+CO$_{2}$ Ice Abundance in Exocomets and Solar System Comets}",
      journal = {\apj},
         year = 2017,
        month = jun,
       volume = {842},
       number = {1},
          eid = {9},
        pages = {9},
          doi = {10.3847/1538-4357/aa71b4},
archivePrefix = {arXiv},
       eprint = {1705.05868},
 primaryClass = {astro-ph.EP},
       adsurl = {https://ui.adsabs.harvard.edu/abs/2017ApJ...842....9M}
}

@ARTICLE{Kral2017,
       author = {{Kral}, Q. and {Matr{\`a}}, L. and {Wyatt}, M. C. and {Kennedy}, G. M.},
        title = "{Predictions for the secondary CO, C and O gas content of debris discs from the destruction of volatile-rich planetesimals}",
      journal = {\mnras},
         year = "2017",
       volume = {469},
        pages = {521-550},
          doi = {10.1093/mnras/stx730}
}

@ARTICLE{Landolt1973,
       author = {{Landolt}, A. U.},
        title = "{UBV photoelectric sequences in the celestial equatorial Selected Areas 92-115}",
      journal = {Astronomical Journal},
         year = 1973,
       volume = {78},
        pages = {959-981},
          doi = {10.1086/111503}
}

@ARTICLE{2018NatAs...2..349E,
       author = {{Ehgamberdiev}, Shuhrat},
        title = "{Modern astronomy at the Maidanak observatory in Uzbekistan}",
      journal = {Nature Astronomy},
         year = 2018,
        month = may,
       volume = {2},
        pages = {349-351},
          doi = {10.1038/s41550-018-0459-3},
       adsurl = {https://ui.adsabs.harvard.edu/abs/2018NatAs...2..349E}
}

@article{MaidanakCCD,
author = {Im, Myungshin and Go, Jong-Wan and Cho, Yun-Seok and Choi, Changsu and Jeon, Yiseul and Lee, In-Deok},
  title     = {SEOUL NATIONAL UNIVERSITY 4K×4K CAMERA (SNUCAM) FOR MAIDANAK OBSERVATORY},
  journal   = {Journal of The Korean Astronomical Society},
  publisher = {한국천문학회},
  volume    = {43},
  number    = {3},
  pages     = {75-93},
  year      = 2010,
  month     = 06,
        doi = {10.5303/PKAS.2008.23.1.001}
}

@article{oke1990faint,
  title={Faint spectrophotometric standard stars},
  author={Oke, JB},
  journal={Astronomical Journal (ISSN 0004-6256), vol. 99, May 1990, p. 1621-1631. Research supported by the Space Telescope Science Institute.},
  volume={99},
  pages={1621--1631},
  year={1990}
}

@article{landolt1992ubvri,
  title={UBVRI photometric standard stars in the magnitude range 11.5-16.0 around the celestial equator},
  author={Landolt, Arlo U},
  journal={Astronomical Journal (ISSN 0004-6256), vol. 104, no. 1, July 1992, p. 340-371, 436-491. Research supported by Space Telescope Science Institute.},
  volume={104},
  pages={340--371},
  year={1992}
}
\bibliographystyle{aasjournalv7}

\appendix

\section{Observing logs}

\begin{table}[!ht]
    \centering 
    \caption{Log of the photometric observations of 3I/ATLAS obtained at the Maidanak Observatory (Uzbekistan)}
    \label{tab:maidanak_obs}
    \begin{tabular}{lrrlrll}
\hline\hline
\multicolumn{1}{c}{Date} &
\multicolumn{1}{c}{$r_{\mathrm{h}}$} &
\multicolumn{1}{c}{$\alpha$} &
\multicolumn{1}{c}{Filters} &
\multicolumn{1}{c}{$N_{\mathrm{exp}}$} &
\multicolumn{1}{c}{$T_{\mathrm{total}}$} &
\multicolumn{1}{c}{$t_{\mathrm{exp}}$} \\
\multicolumn{1}{c}{(UTC)} &
\multicolumn{1}{c}{\textbf{(au)}} &
\multicolumn{1}{c}{\textbf{(deg)}} &
\multicolumn{1}{c}{} &
\multicolumn{1}{c}{} &
\multicolumn{1}{c}{\textbf{(s)}} &
\multicolumn{1}{c}{\textbf{(s)}} \\
    \hline
    2025-08-01 & 3.492 & 13.8 & $B, R, V$ & 30 & 1800 & 60.0 \\
    2025-08-06 & 3.330 & 15.7 & $B, R, V$ & 28 & 1680 & 60.0 \\
    2025-08-09 & 3.233 & 16.9 & $B, R, V$ & 30 & 1800 & 60.0 \\
    2025-08-12 & 3.137 & 17.9 & $B, R, V$ & 31 & 1860 & 60.0 \\
    2025-08-15 & 3.042 & 18.9 & $B, R, V$ & 30 & 1800 & 60.0 \\
    2025-08-18 & 2.947 & 19.9 & $B, R, V$ & 30 & 1800 & 60.0 \\
    2025-08-21 & 2.853 & 20.7 & $B, R, V$ & 30 & 1800 & 60.0 \\
    2025-08-26 & 2.698 & 21.9 & $B, R, V$ & 30 & 1800 & 60.0 \\
    2025-08-29 & 2.606 & 22.4 & $B, R, V$ & 30 & 1800 & 60.0 \\
    \hline\hline
\end{tabular}

    \tablecomments{$r_{\mathrm{h}}$ is the heliocentric distance of the comet, $\alpha$ is its solar phase angle.}
\end{table}

\begin{table}[!ht]
    \centering
    \caption{Log of the photometric observations of 3I/ATLAS obtained with the Faulkes Telescope North and South (FTN and FTS).}
    \label{tab:lco_obs}
    \begin{tabular}{lrrlrll}
\hline\hline
\multicolumn{1}{c}{Date} &
\multicolumn{1}{c}{$r_{\mathrm{h}}$} &
\multicolumn{1}{c}{$\alpha$} &
\multicolumn{1}{c}{Filters} &
\multicolumn{1}{c}{$N_{\mathrm{exp}}$} &
\multicolumn{1}{c}{$T_{\mathrm{total}}$} &
\multicolumn{1}{c}{$t_{\mathrm{exp}}$} \\
\multicolumn{1}{c}{(UTC)} &
\multicolumn{1}{c}{\textbf{(au)}} &
\multicolumn{1}{c}{\textbf{(deg)}} &
\multicolumn{1}{c}{} &
\multicolumn{1}{c}{} &
\multicolumn{1}{c}{\textbf{(s)}} &
\multicolumn{1}{c}{\textbf{(s)}} \\
\hline
2025-07-26 & 3.688 & 11.3 & $g', i', r', z_s$ & 20 & 1100 & 55.0 \\
2025-07-28 & 3.622 & 12.1 & $g', i', r', z_s$ & 20 & 1100 & 55.0 \\
2025-07-28 & 3.622 & 12.1 & $g', r', z_s$      & 15 & 825  & 55.0 \\
2025-07-29 & 3.590 & 12.5 & $g', i', r', z_s$ & 16 & 880  & 55.0 \\
2025-08-06 & 3.330 & 15.7 & $g', i', r', z_s$ & 5  & 250  & 50.0 \\
2025-08-06 & 3.330 & 15.7 & $g', i', r', z_s$ & 4  & 220  & 55.0 \\
2025-08-07 & 3.298 & 16.1 & $g', i', r', z_s$ & 5  & 275  & 55.0 \\
2025-08-12 & 3.137 & 17.9 & $g', i', r', z_s$ & 4  & 220  & 55.0 \\
2025-08-14 & 3.074 & 18.6 & $g', i', r', z_s$ & 4  & 220  & 55.0 \\
2025-08-15 & 3.042 & 18.9 & $g', i', r', z_s$ & 4  & 220  & 55.0 \\
2025-08-24 & 2.760 & 21.4 & $g', i', r', z_s$ & 24 & 1200 & 50.0 \\
2025-08-25 & 2.729 & 21.7 & $g', i', r', z_s$ & 16 & 880  & 55.0 \\
2025-08-27 & 2.667 & 22.1 & $g', i', r', z_s$ & 16 & 880  & 55.0 \\
2025-08-28 & 2.637 & 22.2 & $g', i', r', z_s$ & 44 & 2200 & 50.0 \\
2025-08-30 & 2.576 & 22.5 & $g', i', r', z_s$ & 24 & 1200 & 50.0 \\
2025-09-02 & 2.485 & 22.9 & $g', i', r', z_s$ & 23 & 1180 & 51.3 \\
2025-09-04 & 2.426 & 23.0 & $g', i', r', z_s$ & 4  & 240  & 60.0 \\
2025-09-04 & 2.426 & 23.0 & $g', i', r', z_s$ & 20 & 1000 & 50.0 \\
2025-09-07 & 2.338 & 23.2 & $g', i', r', z_s$ & 20 & 1000 & 50.0 \\
2025-09-08 & 2.309 & 23.1 & $g', i', r', z_s$ & 20 & 1000 & 50.0 \\
2025-09-11 & 2.222 & 23.0 & $g', i', r', z_s$ & 8  & 480  & 60.0 \\
\hline\hline
\end{tabular}
    
    \tablecomments{$r_{\mathrm{h}}$ is the heliocentric distance of the comet, $\alpha$ is its solar phase angle.}
\end{table}

\begin{table}[!ht]
    \centering
    \caption{Log of the spectroscopic observations of 3I/ATLAS obtained with the SALT and NOT telescopes.}
    \label{tab:spectra_obs}
    \setlength{\tabcolsep}{2.5pt} 
    \begin{tabular}{lccrrrr}
\hline\hline
\multicolumn{1}{c}{Date} &
\multicolumn{1}{c}{Tel} &
\multicolumn{1}{c}{Obj} &
\multicolumn{1}{c}{X} &
\multicolumn{1}{c}{$\lambda$} &
\multicolumn{1}{c}{$R$} &
\multicolumn{1}{c}{$T_{\mathrm{total}}$} \\
\multicolumn{1}{c}{(UTC)} &
\multicolumn{1}{c}{} &
\multicolumn{1}{c}{} &
\multicolumn{1}{c}{} &
\multicolumn{1}{c}{\textbf{(\AA)}} &
\multicolumn{1}{c}{} &
\multicolumn{1}{c}{\textbf{(s)}} \\
\hline
2025-07-15.97 & SALT & Comet & 1.285 & 3600 -- 7200 & 270  & 600 \\
2025-07-15.99 & SALT &   SA  & 1.189 & 3600 -- 7200 & 270  & 100 \\
2025-07-25.99 & NOT  & Comet & 1.622 & 4000 -- 9000 & 360  & 2340 \\
2025-07-26.10 & NOT  &   SA  & 1.151 & 4000 -- 9000 & 360  & 25 \\
2025-07-30.01 & NOT  & Comet & 1.997 & 4000 -- 9000 & 360  & 2340 \\
2025-07-30.03 & NOT  &   SA  & 1.154 & 4000 -- 9000 & 360  & 25 \\
2025-08-29.75 & SALT & Comet & 1.276 & 3600 -- 7200 & 1400 & 1800 \\ 
2025-08-29.79 & SALT &   SA  & 1.346 & 3600 -- 7200 & 1400 & 100 \\ 
2025-08-29.88 & NOT  & Comet & 1.963 & 5600 -- 9500 & 770  & 2340 \\
2025-08-29.89 & NOT  &   SA  & 1.292 & 5600 -- 9500 & 770  & 25 \\
2025-09-03.87 & NOT  & Comet & 2.257 & 3600 -- 5800 & 490  & 2340 \\
2025-09-03.89 & NOT  &   SA  & 1.246 & 3600 -- 5800 & 490  & 25 \\
\hline\hline
\end{tabular}

    \tablecomments{In the table \textit{Date} refers to the middle of observation, \textit{Tel} denotes the telescope used, \textit{Obj} denotes the observed object (where \textit{Comet} means 3I/ATLAS and \textit{SA} is the Solar analog star BD-004074 \citealt{landolt1992ubvri}), \textit{X} is the airmass, $\lambda$ gives the spectral range, $R$ is the spectral resolving power, and $T_{\mathrm{total}}$ is the total observing time after stacking multiple individual exposures.}
\end{table}

\begin{table}
    \centering
    \caption{Faulkes Telescopes photometry of 3I/ATLAS measured within a projected 
cometocentric radius of 10,000~km. Date and JD give the UTC observing date 
and mean Julian date, respectively, while $D_{\odot}$ is the comet heliocentric 
distance in AU. The $g'$, $r'$, $i'$, and $z_s$ columns contain apparent 
magnitudes and their associated $1\sigma$ uncertainties.}
    \begin{tabular}{lcccccccccc}
\hline\hline
Date & JD & $D_{\odot}$ & $g'$ & $\delta g'$ & $r'$ & $\delta r'$ & $i'$ & $\delta i'$ & $z_s$ & $\delta z_s$ \\
& (UTC) & \textbf{(au)} & \textbf{(mag)} & \textbf{(mag)} & \textbf{(mag)} & \textbf{(mag)} & \textbf{(mag)} & \textbf{(mag)} & \textbf{(mag)} & \textbf{(mag)} \\
\hline
20250721 & 2460877.917 & 3.838 & 18.263 & 0.016 & 17.581 & 0.011 & 17.392 & 0.031 & 17.275 & 0.011 \\
20250727 & 2460884.876 & 3.610 & 17.987 & 0.006 & 17.342 & 0.005 & 17.081 & 0.007 & 17.010 & 0.008 \\
20250728 & 2460885.796 & 3.580 & 17.960 & 0.005 & 17.313 & 0.005 & 17.047 & 0.007 & 16.958 & 0.005 \\
20250806 & 2460893.966 & 3.315 & 17.643 & 0.017 & 16.955 & 0.010 & 16.749 & 0.007 & 16.652 & 0.010 \\
20250807 & 2460895.868 & 3.254 & 17.553 & 0.015 & 16.923 & 0.008 & 16.668 & 0.009 & 16.557 & 0.008 \\
20250812 & 2460900.749 & 3.098 & 17.382 & 0.005 & 16.722 & 0.003 & 16.461 & 0.006 & 16.373 & 0.005 \\
20250814 & 2460902.766 & 3.034 & 17.319 & 0.003 & 16.649 & 0.003 & 16.369 & 0.006 & 16.284 & 0.007 \\
20250815 & 2460903.742 & 3.003 & 17.269 & 0.004 & 16.608 & 0.005 & 16.340 & 0.004 & 16.252 & 0.005 \\
20250823 & 2460911.742 & 2.752 & 16.950 & 0.004 & 16.299 & 0.005 & 16.027 & 0.004 & 15.941 & 0.007 \\
20250824 & 2460912.745 & 2.721 & 16.904 & 0.005 & 16.254 & 0.004 & 15.995 & 0.005 & 15.906 & 0.004 \\
20250826 & 2460914.736 & 2.660 & 16.825 & 0.005 & 16.191 & 0.004 & 15.911 & 0.004 & 15.821 & 0.006 \\
20250828 & 2460915.917 & 2.624 & 16.793 & 0.005 & 16.124 & 0.005 & 15.866 & 0.009 & 15.780 & 0.004 \\
20250830 & 2460917.942 & 2.562 & 16.761 & 0.033 & 16.024 & 0.018 & 15.774 & 0.008 & 15.687 & 0.012 \\
20250902 & 2460920.895 & 2.474 & 16.605 & 0.017 & 15.936 & 0.013 & 15.674 & 0.011 & 15.568 & 0.018 \\
20250903 & 2460922.751 & 2.418 & 16.515 & 0.006 & 15.867 & 0.005 & 15.589 & 0.009 & 15.523 & 0.008 \\
20250904 & 2460922.927 & 2.413 & 16.503 & 0.008 & 15.852 & 0.008 & 15.605 & 0.004 & 15.507 & 0.003 \\
20250906 & 2460925.731 & 2.331 & 16.383 & 0.020 & 15.747 & 0.005 & 15.493 & 0.004 & 15.405 & 0.009 \\
20250907 & 2460926.731 & 2.302 & 16.325 & 0.017 & 15.733 & 0.006 & 15.449 & 0.013 & 15.363 & 0.007 \\
20250910 & 2460929.725 & 2.216 & 16.214 & 0.005 & 15.594 & 0.004 & 15.342 & 0.007 & 15.252 & 0.006 \\
\hline
\end{tabular}

    \label{tab:lco_photometry}
\end{table}

\begin{table}
    \centering
    \caption{Maidanak photometry of 3I/ATLAS measured within a projected 
cometocentric radius of 10,000~km. Date and JD give the UTC observing date 
and mean Julian date, respectively, while $D_{\odot}$ is the comet heliocentric 
distance in AU. The $B$, $V$, and $R$ columns contain apparent magnitudes 
and their associated $1\sigma$ uncertainties.}
    \begin{tabular}{lcccccccc}
\hline\hline
Date & JD & $D_{\odot}$ & $B$ & $\delta B$ & $V$ & $\delta V$ & $R$ & $\delta R$ \\
\hline
20250801 & 2460889.189 & 3.470 & 17.503 & 0.017 & 16.665 & 0.013 & 16.045 & 0.009 \\
20250806 & 2460894.190 & 3.308 & 17.338 & 0.047 & 16.405 & 0.020 & 15.848 & 0.015 \\
20250809 & 2460897.184 & 3.212 & 17.258 & 0.061 & 16.243 & 0.021 & 15.722 & 0.020 \\
20250812 & 2460900.174 & 3.116 & 17.058 & 0.013 & 16.174 & 0.006 & 15.626 & 0.005 \\
20250815 & 2460903.172 & 3.021 & 16.933 & 0.010 & 16.048 & 0.006 & 15.513 & 0.006 \\
20250821 & 2460909.156 & 2.833 & 16.671 & 0.022 & 15.828 & 0.009 & 15.281 & 0.010 \\
20250826 & 2460914.153 & 2.678 & 16.475 & 0.032 & 15.626 & 0.019 & 15.093 & 0.026 \\
20250829 & 2460917.146 & 2.586 & 16.362 & 0.014 & 15.501 & 0.020 & 14.985 & 0.032 \\
\hline
\end{tabular}

    \label{tab:maidanak_photometry}
\end{table}



\end{document}